\documentclass[epj]{svjour}
\usepackage{amsmath}
\usepackage[switch]{lineno} 
\usepackage{xspace} 
\usepackage{hyperref} 
\usepackage{color} 
\usepackage{array}
\newcolumntype{P}[1]{>{\centering\arraybackslash}p{#1}}
\usepackage{subcaption}
\usepackage{graphicx}
\usepackage{doi}
\usepackage{multirow}

\def\u{$^{238}$U\xspace}
\def\th{$^{232}$Th\xspace}
\def\k{$^{40}$K\xspace}
\def\kqu{$^{41}$K\xspace}
\def\np{$^{239}$Np\xspace}
\def\pa{$^{233}$Pa\xspace}
\def\kqd{$^{42}$K\xspace}

\def\cocn{$^{59}$Co\xspace}
\def\cosz{$^{60}$Co\xspace}

\def\auuns{$^{197}$Au\xspace}
\def\auuno{$^{198}$Au\xspace}

\def\alds{$^{27}$Al\xspace}
\def\nadq{$^{24}$Na\xspace}

\def\nalfa{(n,$\alpha$)\xspace}
\def\ngamma{(n,$\gamma$)\xspace}
\begin{document}
%
\title{Radiopurity screening of materials for rare event searches by neutron activation at the TRIGA reactor of Pavia}
\author{G. Baccolo\inst{1,2,}\thanks{\emph{Present address:} Dipartimento di Scienze, Universit\`a degli Studi di Roma Tre, Rome 00146, Italy}%
\and A. Barresi\inst{1,3}
\and D. Chiesa\inst{1,3,}\thanks{\emph{Corresponding author:} davide.chiesa@mib.infn.it}
\and M. Nastasi\inst{1,3}
\and E. Previtali\inst{1,3}
\and M. Sisti\inst{1,}\thanks{\emph{Corresponding author:} monica.sisti@mib.infn.it}
}                     
%
%
\institute{INFN -- Sezione di Milano Bicocca, 20126 Milan, Italy
\and Dipartimento di Scienze dell'Ambiente e della Terra - DISAT, Universit\`a degli Studi di Milano-Bicocca, 20126 Milan, Italy
\and Dipartimento di Fisica ``Giuseppe Occhialini", Universit\`a degli Studi di Milano-Bicocca, 20126 Milan, Italy
}
\date{Received: date / Revised version: date}
%
\abstract{
In the framework of physics experiments searching for rare events, the selection of extremely radiopure materials is a challenging task, as the signal of interest is often hidden by instrumental background.
Neutron activation is a powerful technique to measure trace contaminants with high sensitivity but, to be properly applied, it requires a good characterization of neutron irradiation and $\gamma$-spectroscopy facilities.
This paper presents the state-of-the-art workflow adopted by the Radioactivity Laboratory of the University of Milano-Bicocca for radiopurity screening of materials by neutron activation performed at the TRIGA reactor in Pavia. The ultimate sensitivity of the described workflow, in the absence of interfering activation products and without the application of radiochemical and/or active background rejection techniques, is $<10^{-14}$ g/g for $^{40}$K ($<10^{-10}$ g/g for elemental K), $<10^{-12}$ g/g for $^{238}$U and of the order of $10^{-12}$ g/g for $^{232}$Th contaminations.
Further details are here provided to address systematic uncertainties related to neutron irradiation that may bias results. To this aim, a dedicated neutron activation campaign was performed and the data were analyzed exploiting a Monte Carlo simulation model of the reactor and applying an unfolding technique to obtain a comprehensive characterization of the neutron flux in typical irradiation configurations.
The results of this work provide a valuable benchmark for the application of neutron activation in future radiopurity screening campaigns.
\PACS{
      {PACS-key}{discribing text of that key}   \and
      {PACS-key}{discribing text of that key}
     } 
} 
\authorrunning{G. Baccolo, et al}
\titlerunning{Radiopurity screening of materials for rare event searches by neutron activation}
\maketitle
\section{Introduction}
\label{sec:intro}
Experiments searching for rare particle physics events demand increasingly higher levels of radiopurity for the materials that make up the detectors~\cite{heusser:1995,Laubenstein:2020rbe}. The ultimate experimental sensitivity critically depends on the level of residual radioactive background in the energy region of interest, which could hamper the detection of the very faint signals being sought. 
Besides locating these experiments in deep underground laboratories to reduce cosmogenic-induced events~\cite{Laubenstein:2004}, surrounding the detectors with massive shielding to attenuate environmental background~\cite{Smith:2008}, and employing particle identification methods to discard spurious energy depositions~\cite{Budjas:2009zu}, it is crucial to construct the detector using materials with the lowest possible concentration of radioactive contaminants.
In this respect, the assay techniques also require increasingly higher levels of sensitivity to evaluate the radiopurity of materials.

Typical contaminants include primordial radionuclides such as \k, \th, $^{235}$U and \u, as well as their decay chain daughters (for Th and U isotopes). Depending on the materials, a few cosmogenic and artificial long-lived radionuclides, such as $^{60}$Co and $^{137}$Cs, may also be of concern.
The most widely used screening technique to study bulk contamination of materials is $\gamma$-spectroscopy~\cite{Laubenstein:2017yjj}, which has the significant advantage of being sensitive to most nuclides in natural decay chains and to many cosmogenic and anthropogenic radioisotopes. Alpha spectroscopy is typically used to evaluate the presence of \th, \u, and their $\alpha$-decaying daughters (particularly radon progenies) on material surfaces~\cite{Bunker:2020sxw}.
Complementary bulk contamination screening techniques include mass concentration measurement methods like Inductively Coupled Plasma Mass Spectrometry (ICP-MS) and Neutron Activation Analysis (NAA)~\cite{Arpesella:2001iz,Nisi:2009}. Although these techniques are limited to detecting only the parent nuclides, such as \th and \u, without the ability to assess the concentrations of their daughters, they can achieve much higher analytical sensitivities compared to $\gamma$-spectroscopy.

NAA is an established technique for analyzing the elemental composition of various samples~\cite{Greenberg:2011}. In the context of radioactivity measurements, NAA offers several advantages. It is highly versatile, allowing for the analysis of a wide range of samples (powdered, solid, liquid) without the need for dedicated preparation procedures. Additionally, it is highly sensitive to primordial radionuclides, enabling the quantification of contaminations down to $10^{-12}$~g/g, or even lower in some specific cases or when coupled with radiochemistry~\cite{Goldbrunner:1998bq}. Another significant advantage is its low contamination risk: after neutron irradiation, the possibility of contaminating the samples is almost negligible since the analytes of interest are artificial radionuclides not present in the environment. For these reasons, NAA has been widely used as a tool to evaluate radiopurity in various samples across many experiments. In the context of rare event physics experiments, typical samples that have been radiologically screened through NAA include metals, plastic components, electronics, reagents, and inorganic crystals~\cite{Arpesella:2001iz,Alessandrello:1998ats,Buehler:1995kc,Djurcic:2002uu,Leonard:2007uv,Salvini:2006,Alduino:2017qet}.
However, NAA also has some disadvantages. Besides being able to quantify only the concentration of primordial radionuclides, it requires access to a nuclear reactor, with measurements being conducted by staff trained to handle radioactive materials. Another critical issue with NAA is the correct assessment of uncertainties, particularly those related to systematic errors.

The aim of the present work is to present the state-of-the-art workflow adopted at the Radioactivity Laboratory of the University of Milano-Bicocca with respect to the radiopurity screening of materials to be used in rare event physics experiments through NAA. The most important practical issues which influence the measurement process will be discussed, from the planning of the irradiation to the analysis of $\gamma$-spectra. Particular attention will be given to the management of uncertainties and of systematic errors.

\section{Experimental methodology}
\label{sec:method}
The main step of NAA is neutron irradiation, usually at a research reactor. The contaminants to be measured are targets of neutron-induced reactions, which lead to the production of radionuclides. 
Subsequently, $\gamma$-spectroscopy measurements are performed to measure the induced activities and determine the concentrations of contaminants in the sample.
High-Purity Germanium (HPGe) detectors are the most widely used $\gamma$-spectrometers, because of their excellent energy resolution.

The activation reactions usually exploited for natural contaminant concentration measurements are radiative $(n,\gamma)$ captures on \u, \th and \kqu, generating radionuclides according to the following sequences:
\begin{align*}
    & ^{238}\text{U} \xrightarrow{(n,\gamma)} \, ^{239}\text{U} \xrightarrow[23.45\text{ m}]{\beta^-} \, ^{239}\text{Np} \xrightarrow[2.356\text{ d}]{\beta^-} \, ^{239}\text{Pu}\\
    & ^{232}\text{Th} \xrightarrow{(n,\gamma)} \, ^{233}\text{Th} \xrightarrow[22.3\text{ m}]{\beta^-} \, ^{233}\text{Pa} \xrightarrow[26.97 \text{ d}]{\beta^-} \, ^{233}\text{U}\\
    & ^{41}\text{K} \xrightarrow{(n,\gamma)} \, ^{42}\text{K} \xrightarrow[12.36\text{ h}]{\beta^-}  \, ^{42}\text{Ca}
\end{align*}
Given the relatively short half-lives of $^{239}$U and $^{233}$Th, their daughters \np and \pa are most conveniently used to measure \u and \th concentrations.

The key parameter of NAA is the activation rate ($R$) of each reaction, because it is directly proportional to the number of target isotopes ($\mathcal{N}$) that are used to determine the contaminant concentration in the sample:
\begin{equation}
\label{Eq:ActRate}
R = \mathcal{N} \int \sigma(E) \, \varphi(E) \, dE
\end{equation}
where $\sigma(E)$ is the activation cross section, and $\varphi(E)$ is the differential distribution of the neutron flux ($\phi$) as a function of neutron energy ($\varphi(E) \equiv d\phi / dE$).  

The activities of the different radionuclides are directly proportional to the corresponding activation rates and, after a certain \textit{cooling} time ($t_\text{c}$) between the end of irradiation and the beginning of the $\gamma$-spectroscopy measurement, are equal to\footnote{For $t$ much greater than the half-lives of $^{239}\text{U}$ and $^{233}\text{Th}$, Eq.~\ref{Eq:Activity} can be used to calculate the activities of \np and \pa with very good approximation.}:
\begin{equation}
A_0 = R \, (1-e^{-\lambda t_{\text{irr}}}) \, e^{-\lambda t_\text{c}}
\label{Eq:Activity}
\end{equation}
where $t_{\text{irr}}$ is the irradiation time and $\lambda \equiv \ln{2} / T_{1/2}$ is the decay constant of the isotope under study. It is worth noting that the maximum activity that can be reached (when $t_{\text{irr}} \gg T_{1/2}$) is the activation rate $R$, which is thus also referred to as \textit{saturation activity}. 
At this point, to take into account the decrease of activity during $\gamma$-spectroscopy, it is convenient to introduce the number of decays ($n_\text{dec}$) that are expected to occur while measuring for a time $t_\text{m}$:
\begin{equation}
n_\text{dec} = \int_0^{t_\text{m}} A_0 \, e^{-\lambda t} dt=A_0 \dfrac{1-e^{-\lambda t_\text{m}}}{\lambda}
\label{Eq:nDec}
\end{equation}
In this way, we get the equation of $R$ as a function of $n_\text{dec}$ and of the experimental times $t_\text{irr}$, $t_\text{c}$ and $t_\text{m}$:  
\begin{equation}
R = \dfrac{n_\text{dec} \, \lambda}{(1-e^{-\lambda t_\text{irr}})e^{-\lambda t_\text{c}}(1-e^{-\lambda t_\text{m}})}
\label{Eq:R_vs_nDec}
\end{equation}

\begin{table}
\begin{tabular}{llll}
\hline\noalign{\smallskip}
Isotope & Half-life & Energy (keV) & $I_{\gamma}$ (\%)  \\
\noalign{\smallskip}\hline\noalign{\smallskip}
\kqd & 12.36 h & 1524.7 & 18.1\\
\np & 2.356 d & 106.1 & 25.3\\
\pa & 26.97 d & 311.9 & 38.2\\
\noalign{\smallskip}\hline
\end{tabular}
\caption{List of the main $\gamma$-rays emitted by the \u, \th and \kqu activation products. The intensity $I_{\gamma}$ reported in the last column is the probability of emission of that $\gamma$-ray for each decay.}
\label{Tab:gammalines}       
\end{table}

In $\gamma$-spectroscopy measurements, the characteristic $\gamma$-lines of the activated isotopes (Tab.~\ref{Tab:gammalines}) are analyzed to determine $n_\text{dec}$, and thus $R$, from the net number of counts ($C$) recorded in the corresponding peaks.
The coefficient of proportionality between $C$ and $n_\text{dec}$ is the product of the full-energy detection efficiency ($\varepsilon_{\gamma}$) and the emission probability ($I_{\gamma}$) of the $\gamma$-ray in question:
\begin{equation}
    C = \varepsilon_{\gamma} \, I_{\gamma} n_\text{dec}
\label{Eq:C_vs_nDec}
\end{equation}
While $I_{\gamma}$ data are usually known with good precision, the efficiency $\varepsilon_{\gamma}$ is a delicate parameter that depends on:
\begin{itemize}
    \item the geometry and materials of both the detector and the sample;
    \item the position of the sample with respect to the detector;
    \item the $\gamma$-ray energy; 
    \item the decay scheme of the radioisotope, especially in the case of $\gamma$ cascades that cannot be resolved in time, giving rise to the so-called \textit{sum peaks}.
\end{itemize}
Therefore, even if it is possible to build efficiency curves as a function of $\gamma$-ray energy using calibrated multi-$\gamma$ sources, Monte Carlo methods should be preferred when dealing with various samples having different geometries and materials, because they allow to achieve higher accuracy in $\varepsilon_{\gamma}$ evaluation.

Once $R$ is determined from the experimental data, Eq.~\ref{Eq:ActRate} must be used to get $\mathcal{N}$ and, thus, the contaminant concentration.
In general, it is not easy to determine the neutron flux intensity and energy spectrum at irradiation facilities with good accuracy. For example, at a nuclear reactor, the neutron flux depends on the operating conditions (control rod positions, fuel and moderator temperatures, fuel burnup), but also on the sample position inside the facility. Besides this, as shown in Sect.~\ref{sec:systematics}, the sample itself can introduce a local distortion of the flux because of neutron moderation, scattering, and/or self-shielding effects.

The most popular approach to overcome this issue is to irradiate the samples together with the so-called \textit{standards}, i.e. samples containing known amounts of the elements of interest. 
If it is possible to assume that the sample and its \textit{standard} are co-irradiated with the same neutron flux, the analysis can proceed with a remarkable simplification.
Indeed, it is possible to write the following equation:
\begin{equation}
\dfrac{\mathcal{N}}{\mathcal{N}^{std}}=
\dfrac{R}{R^{std}}=
\dfrac{C}{C^{std}} \, \dfrac{\varepsilon_{\gamma}^{std}}{\varepsilon_{\gamma}} \, \dfrac{e^{-\lambda t_\text{c}^{std}} \left( 1-e^{-\lambda t_\text{m}^{std}} \right) }{e^{-\lambda t_\text{c}} \left(1-e^{-\lambda t_\text{m}} \right)}
\label{Eq:STD_method}
\end{equation}
where the superscript $std$ is used to label the parameters of the \textit{standard}.
It is worth noting here that if the \textit{standard} has the same geometry and materials of the sample, the efficiency ratio $\varepsilon_{\gamma} / \varepsilon_{\gamma}^{std}$ can be assumed equal to 1, otherwise a Monte Carlo simulation is usually needed to determine this corrective factor with good accuracy.

When NAA is not limited to radiopurity assessment, it is not possible or it is not practically convenient to irradiate and measure \textit{standard} samples for all the elements of interest.
In such cases, the so-called $k_0$ method~\cite{Simonits197531} allows to determine the masses of the elements through a single flux monitor (or \textit{comparator}), generally made of gold. This method exploits constant factors, called $k_0$, that were accurately measured for the $\gamma$-rays of most isotopes produced by neutron activation~\cite{DeCorte200347}.
The $k_0$ factor is indeed a nuclear constant that corresponds to the ratio between the $\gamma$-rays per unit mass emitted by the element in question $x$ and by the comparator $c$, when the activation is induced only by thermal neutrons:
\begin{equation}
    k_{0,x}=\dfrac{I_{\gamma,x} \, (\theta_x / M_x) \, \sigma_{0,x}}{I_{\gamma,c} \, (\theta_c / M_c) \, \sigma_{0,c}}
\end{equation}
where $\theta$ is the isotopic abundance of the target isotope, $M$ the atomic mass, and $\sigma_{0}$ the thermal activation cross section.
In order to get accurate results with the $k_0$ method when the irradiation is performed with both thermal and epithermal neutrons, it is necessary to:
\begin{itemize}
    \item know the ratio of the resonance integral to the thermal activation cross section (which is usually tabulated together with $k_0$ values);
    \item perform a careful calibration of the irradiation facility to determine the ratio of thermal to epithermal neutron flux and the shape parameter $\alpha$ used to describe the flux spectrum ($\propto 1/E^{1+\alpha}$) in the intermediate range;
    \item know the efficiency ratio for the detection of $\gamma$-rays emitted by the sample and by the comparator, respectively;
    \item take into account effects related to neutron flux inhomogeneity due to neutron scattering or absorption by the sample itself.
\end{itemize}
The success of the $k_0$ standardization method is related to the fact that the experimenter, once the irradiation and $\gamma$-spectroscopy facilities are characterized, needs only to count the number of $\gamma$-rays detected at the full-energy peaks, and to take note of irradiation, cooling and measurement times.
More details about the application of $k_0$ method can be found in~\cite{Review_k0_2012}.

Finally, if the $k_0$ value is not tabulated for a specific isotope of interest and it is not possible to prepare a \textit{standard} for that element (e.g. in the case of a noble gas), its concentration can be determined by calculating the \textit{effective cross section} $\sigma_\text{eff}$.
This parameter is defined as:
\begin{equation}
\label{Eq:XSeff}
    \sigma_\text{eff}=\dfrac{\int \sigma(E) \, \varphi(E) \, dE}{\int \varphi(E) \, dE}
\end{equation}
and corresponds to the average cross section weighted on the flux spectrum.
By calling $\Phi \equiv \int \varphi(E) \, dE$ the total flux intensity, Eq.~\ref{Eq:ActRate} can be re-written as:
\begin{equation}
R=\mathcal{N} \,  \sigma_\text{eff} \, \Phi
\end{equation}
When a comparator $c$ (or, equivalently, a standard) is co-irradiated with the sample, it is possible to simplify the flux intensity $\Phi$, obtaining the following identity:
\begin{equation}
    \dfrac{R}{R_c} = \dfrac{\mathcal{N}}{\mathcal{N}_c} \dfrac{\sigma_\text{eff}}{\sigma_{\text{eff},c}}
\end{equation}
It is worth noting that a dependence on the flux spectrum $\varphi(E)$ is embedded in the effective cross section terms. 
Recalling that $\mathcal{N}= \theta \, N_\text{Av} \, m / M$ ($m$ being the element mass), Eq.~\ref{Eq:STD_method} becomes:
\begin{equation}
\dfrac{m}{m_c}=
\dfrac{M \, \theta_c \, \mathcal{N} }{M_c \, \theta \, \mathcal{N}_c } =\dfrac{M \, \theta_c}{M_c \, \theta} \dfrac{\sigma_{\text{eff},c}}{\sigma_\text{eff}} \dfrac{R}{R_c}
\end{equation}
where the ratio $R/R_c$ can be determined from the $\gamma$-spectroscopy measurements using Eq.~\ref{Eq:R_vs_nDec} and \ref{Eq:C_vs_nDec}.
As mentioned earlier, this method can be used to analyze any activation reaction, provided that the neutron spectrum $\varphi(E)$, is known with sufficient accuracy and level of detail -- which is often not the case. 
In this context, as will be discussed in Sect.~\ref{sec:systematics}, a Monte Carlo simulation of reactor neutronics allows to access such information and, more generally, it represents a powerful tool for the analysis of self-shielding effects of irradiated samples and of spatial gradients.

\subsection{NAA sensitivity}
\label{sec:NAAsensitivity}

The sensitivity of NAA essentially depends on the capability of detecting the activated radionuclides. This, in turn, implies an optimised planning of the activation experiment by carefully choosing the neutron irradiation parameters and analyzing the characteristics of the isotope under investigation. In particular, the isotopes that are best suited for NAA should have a sufficiently high neutron activation cross section, and the resulting activated radionuclides should have a decay half-life of the same order of magnitude as the duration of typical $\gamma$-spectroscopy measurements (i.e., from a few hours up to few weeks) as well as a high probability of $\gamma$ emission. 
All of these elements, not least the energy of the emitted $\gamma$ and the sample geometry, will affect the ability to discriminate the expected $\gamma$ peak from the HPGe background.
For a given isotope, according to Eq.~\ref{Eq:ActRate} and Eq.~\ref{Eq:Activity}, the experimental parameters that can be acted upon to maximize the activity $A_0$ at the beginning of the $\gamma$-spectroscopy measurement are:
\begin{itemize}
    \item the irradiation time $t_{\text{irr}}$;
    \item the \textit{cooling} time $t_{\text{c}}$;
    \item the number of target isotopes $\mathcal{N}$, which scales proportionally to the sample mass (or surface);
    \item the neutron flux intensity and energy spectrum (which vary according to the irradiation position within the nuclear reactor).
\end{itemize}
Typically, one has to find the best compromise that allows to respect the reactor constraints in terms of duty cycle, neutron flux, available volume in the irradiation facilities, and radiation protection protocols. 
In this respect, further constraints may arise from the sample itself because one has to guarantee that the activation of its matrix and/or the radiolysis of its molecules will not prevent it from being handled and measured after irradiation. Also the stability of the sample container during irradiation as well as its radiopurity are important aspects to consider, to avoid the risk of loosing or contaminating the sample.

After neutron irradiation, the samples are extracted from the reactor and prepared for $\gamma$-spectroscopy, typically by means of HPGe detectors. In low-background laboratories, HPGe detectors are optimised against background in the best possible configuration that allows for maximum sensitivity.
Usually, the detectors are surrounded by thick layers of composite passive shields and, possibly, a muon veto system and/or a Rn abatement system may complete the setup~\cite{heusser:1995}. In this regard, it is important to highlight that, in NAA, the effective background rate of a HPGe detector can be increased by the irradiated sample itself (\emph{correlated} background). This happens when the sample matrix contains \textit{interfering} elements that are activated because of their higher neutron cross sections and/or concentrations, as compared to the searched contaminants. As a result, a non-negligible background may be produced in the energy regions of the $\gamma$-lines of interest. This correlated background can significantly impair the NAA sensitivity if the interfering isotopes have decay times similar to or longer than the half-lives of the searched activation products. In such cases, possible solutions are the selection of the decay patterns of interest by means of $\beta-\gamma$~\cite{gespark,BARRESI2025170322} or $\gamma-\gamma$ coincident systems (depending on the nuclear decay of the activated isotope), the discrimination of Compton events by pulse shape analysis~\cite{BeGeCompton}, or the use of radiochemical treatments when the sample matrix allows it.

The aim of the analysis of the acquired $\gamma$ spectra is to get the number of counts $C$ (see Eq.~\ref{Eq:C_vs_nDec}) recorded in the sought-after peaks. 
If no peak is observed at the predicted energy, one can calculate the \textit{detection limit} for the signal counts $C^*$, applying the statistical method outlined by Currie in 1968~\cite{Currie1968}. This limit corresponds to the \textit{true} signal that has only a 5\% chance of fluctuating below the \textit{critical limit}
\footnote{The \textit{critical limit} is defined as the value such that, if the result of a measurement exceeds it, one can conclude that a non-zero signal has been detected with a 95\% confidence level.}.
Since in $\gamma$-spectroscopy it is usually possible to determine with good precision the background in the signal region, we calculate the Currie's detection limit in the case of ``well-known" background:
\begin{equation}
    C^* =  2.71 + 3.29 \, \sqrt{B \, t_m \, \Delta E}
\label{Eq:Currie}
\end{equation}
where $B$ is the background counting rate per unit energy, $t_m$ the measurement time, and $\Delta E$ is the energy range in which the peak is expected, that we set at 1.2 FWHM (Full Width at Half Maximum) to maximize the signal-to-background ratio~\cite{DEGEER2004151,standard-iso-2010}.
The corresponding upper limit for the number of decays -- to be used for absolute activity calculation -- is given by:
\begin{equation}
    n_\text{dec}^* = \dfrac{C^*}{\varepsilon_{\gamma} \, I_{\gamma} \, f(\Delta E)}
\label{Eq:DetLimit}
\end{equation}
where $f(\Delta E)$ is the fraction of signal events that fall in an energy window $\Delta E$ around the $\gamma$-ray energy ($f(\Delta E)=0.84$ when $\Delta E = 1.2$~FWHM).
Then, by exploiting Eq.~\ref{Eq:R_vs_nDec}, one can set an upper limit on the activation rate $R$, which can be eventually converted into an upper limit for the contaminant concentration within the sample, using one of the methods outlined in the previous section. 

\section{Measurement and analysis workflow}
\label{sec:workflow}
In the last years, at Milano-Bicocca, we optimised the procedures to detect trace concentrations of natural contaminants ($^{40}$K, $^{232}$Th, $^{238}$U) in different materials, by means of the neutron activation reactions listed in Sect.~\ref{sec:method}

In the following we briefly discuss the sequence of steps that we usually follow during our irradiation campaigns.

\subsection{Sample preparation}
The presence of natural contaminants is ubiquitous in our environment, so the risk of contaminating a sample when aiming for ppt-level\footnote{A contaminant concentration within a matrix is usually expressed in mBq/kg or g/g, i.e. grams of contaminant per gram of material, and its sub-multiples ppm\,$=10^{-6}$\,g/g, ppb\,$=10^{-9}$\,g/g, and ppt\,$=10^{-12}$\,g/g). For convenience, we report the conversions between the two units: 1\,mBq of $^{232}$Th activity per kg of material is equal to $2.5 \times 10^{-10}$\,g/g (or 0.25\,ppb) of $^{232}$Th mass concentration in that material. Similarly, 1\,mBq/kg of $^{238}$U corresponds to $8.1 \times 10^{-11}$\,g/g (or 81\,ppt) of $^{238}$U, and 1\,mBq/kg of $^{40}$K corresponds to $3.8 \times 10^{-12}$\,g/g (or 3.8\,ppt) of $^{40}$K (or, equivalently, $3.1 \times 10^{-8}$\,g/g of natural K).} 
sensitivities to their concentration within the sample itself is by no means negligible. Hence, the preparation of a sample and its insertion into the irradiation container must be followed with extreme care.

Usually, the sample volume and shape have constraints due to the space available in the reactor irradiation channels. If shaping of the sample is required, meticulous attention must be put in the choice of the workshop tools to avoid accidental contamination with \k, \th, and \u. For our NAA campaigns, we preferably process our samples by using laser cutting or electrical discharge machining, to reduce at minimum the contact with the material surface. Eventually, a surface cleaning step can be introduced after processing to remove a few hundred microns of potentially contaminated material.
The final step of the sample preparation, i.e. its insertion inside the irradiation container, is usually done inside a clean room (our clean room is a class 10000), after a final wash with MilliQ water to remove any dust. 
Last but not least, sample containers may also be a source of contamination, which is why we most frequently use polyethylene (PE) ones and keep them submerged in a high-purity HNO$_3$ solution for several days before operation.

Irradiation standards are typically prepared inside different irradiation containers from those of the samples, to prevent any potential contamination. Whenever commercially available, we opt for liquid solutions of certified single-element reference materials for each specific contaminant under investigation. For natural contaminants, we rely on liquid solutions containing certified concentrations of K, U, and Th, commonly utilized for high precision ICP-MS  measurements. 
To achieve reproducible geometry and, thus, an accurate evaluation of the HPGe detector geometrical efficiency, calibrated volumes of certified liquid solutions are absorbed onto a small filter paper which we put at the bottom of a polyethylene vial. 

In addition to liquid standards, we usually prepare also solid ones. These consist of certified strands of Al-Co (Co 0.5\% wt.), a few mm in length and 1\,mm in diameter, which are placed in each reactor channel used both for samples and liquid standards during NAA irradiation. This approach allows for cross-verification of results in case of any issues.

\subsection{Irradiation at the TRIGA reactor in Pavia}
\label{sec:triga}
As neutron source, we use the TRIGA Mark II nuclear research reactor of the University of Pavia, not far from the University of Milano-Bicocca. 
It is a pool-type reactor cooled and partly moderated by light water, with a maximum nominal power of 250\,kW. Its cylindrical core ($\sim 46$\,cm of diameter and $\sim 36$\,cm in height) is surrounded by a 30-cm-thick radial graphite reflector and a 10-cm-thick axial graphite reflector at both the top and the bottom of each of the fuel elements. 
It is equipped with several irradiation facilities: among them, the Central Thimble, located at the center of the reactor core, and the Lazy Susan facility -- a specimen rack in a circular well within the radial reflector -- are the most suitable for our NAA analyses. 
More details about the geometry of the reactor and its irradiation facilities can be found in references~\cite{AbsoluteFlux,BayesianSpectrum,FluxDistribution}.
The Lazy Susan facility is particularly attractive since it includes 40 irradiation channels and thus offers high flexibility in the number of samples to be irradiated. Each of these channels has a cylindrical shape with a diameter of $\sim 32$\,mm; the PE irradiation vessels, therefore, also have a cylindrical shape, with an inner diameter of 23\,mm and a useful height of $\sim 80$\,mm. These are the dimensions that limit the size of the samples to be irradiated.
The neutron fluxes within the Pavia TRIGA reactor are about $1.7 \times 10^{13}$\,n/cm$^2$/s in the Central Thimble and about $ 2.2 \times 10^{12}$\,n/cm$^2$/s in the Lazy Susan facility~\cite{BayesianSpectrum}. Typical operation times at full power for this TRIGA reactor are six hours per day. Consequently, our NAA campaigns are structured around six-hour irradiation periods for samples.

\subsection{$\gamma$-spectroscopy measurements at Milano-Bicocca}

After irradiation and once radiation protection measures allow, samples are promptly extracted from the reactor, put in a new container, and sent for $\gamma$-spectroscopy to our Radioactivity Laboratory at the third basement floor of the Department of Physics of the University of Milano-Bicocca. Timing is critical at this stage for achieving optimal sensitivity on $^{42}$K, given its half-life of only 12.36 hours. Typically, we can begin $\gamma$-counting in our laboratory about 4 hours after the reactor shutdown, assuming no unforeseen issues.
However, interfering radioisotopes, such as $^{41}$Ar, $^{24}$Na, and $^{82}$Br in our case, may impede  counting of relatively large samples after such short cooling times. In particular, the decay rate of $^{41}$Ar, which has a relatively short half-life of 109.6 minutes, may significantly increase the dead time of the HPGe detector. In such cases, we usually wait an additional 2 to 6 hours for $^{41}$Ar to decay.

Our Radioactivity Laboratory is equipped with five high-resolution HPGe detectors, 
with different relative efficiencies and background shieldings, along with one $\gamma - \gamma$  and one $\beta - \gamma$ low-background coincidence systems. 
Neutron irradiated samples for NAA analysis are usually counted on two of these detectors, named GeGEM and BEGe. 
GeGEM is a p-type coaxial HPGe detector by Ortec, with a relative efficiency of 30\%, and a crystal mass of 815\,g; it is surrounded by a composite shielding of 5\,cm of copper and 12\,cm of lead. 
BEGe is a planar-type broad energy HPGe by Canberra, with a relative efficiency of 50\%, and a crystal mass of 802\,g; it is surrounded by a composite shielding of 10\,cm of copper and 15\,cm of lead. 
Both detectors have an energy resolution of 1.9\,keV at the 1.33\,MeV line of $^{60}$Co.
The residual background rates of the two detectors in the main energy regions of the activation products of $^{41}$K, $^{232}$Th, and $^{238}$U are reported in Table \ref{Tab:hpge-bkg}. It is worth noting that these rates relate only to the constant background associated with the HPGe experimental setup and do not include the correlated time-dependent background originating from the sample after neutron irradiation.

\begin{table}
\begin{tabular}{cccc}
\hline\noalign{\smallskip}
\multirow{ 2}{*}{Isotope} & Energy & \multicolumn{2}{c}{Residual background (counts/h)} \\
& range (keV) & GeGEM detector & BEGe detector  \\
\noalign{\smallskip}\hline\noalign{\smallskip}
$^{239}$Np & 101--111 & $61.3 \pm 0.4$ & $37.6 \pm 0.2$\\
$^{233}$Pa & 307--317 & $34.6 \pm 0.3$ & $22.2 \pm 0.2$ \\
$^{42}$K & 1520--1530 & $2.4 \pm 0.1$ & $1.6 \pm 0.1$\\
\noalign{\smallskip}\hline
\end{tabular}
\caption{The residual background rates of the two detectors primarily utilised for NAA $\gamma$-spectroscopy at Milano-Bicocca are provided, along with their associated errors, within an energy range of 10\,keV (corresponding to $\sim$5 FWHM) centered on the most intense $\gamma$ lines from activation products of $^{41}$K, $^{232}$Th, and $^{238}$U. The background measurements lasted 321\,h for GeGEM detector and 651\,h for BEGe detector.}
\label{Tab:hpge-bkg}      
\end{table}

\subsection{Data analysis}
\label{sec:data_analysis}
$\gamma$-spectroscopy measurements for NAA in our laboratory typically range from 12 hours (for \k screening) to 2 to 4 weeks (for \th screening).
The number of counts $C$ in the $\gamma$ peaks from activation products are determined by selecting an energy region at least 2.5\,FWHM wide and conducting a simultaneous fit of the selected peak and the underlying background. The background is typically modeled with a polynomial function of up to the second degree, while the peaks are assumed to follow a Gaussian distribution.

\begin{figure}
\centering
\begin{subfigure}{0.45\textwidth}
    \includegraphics[width=\textwidth]{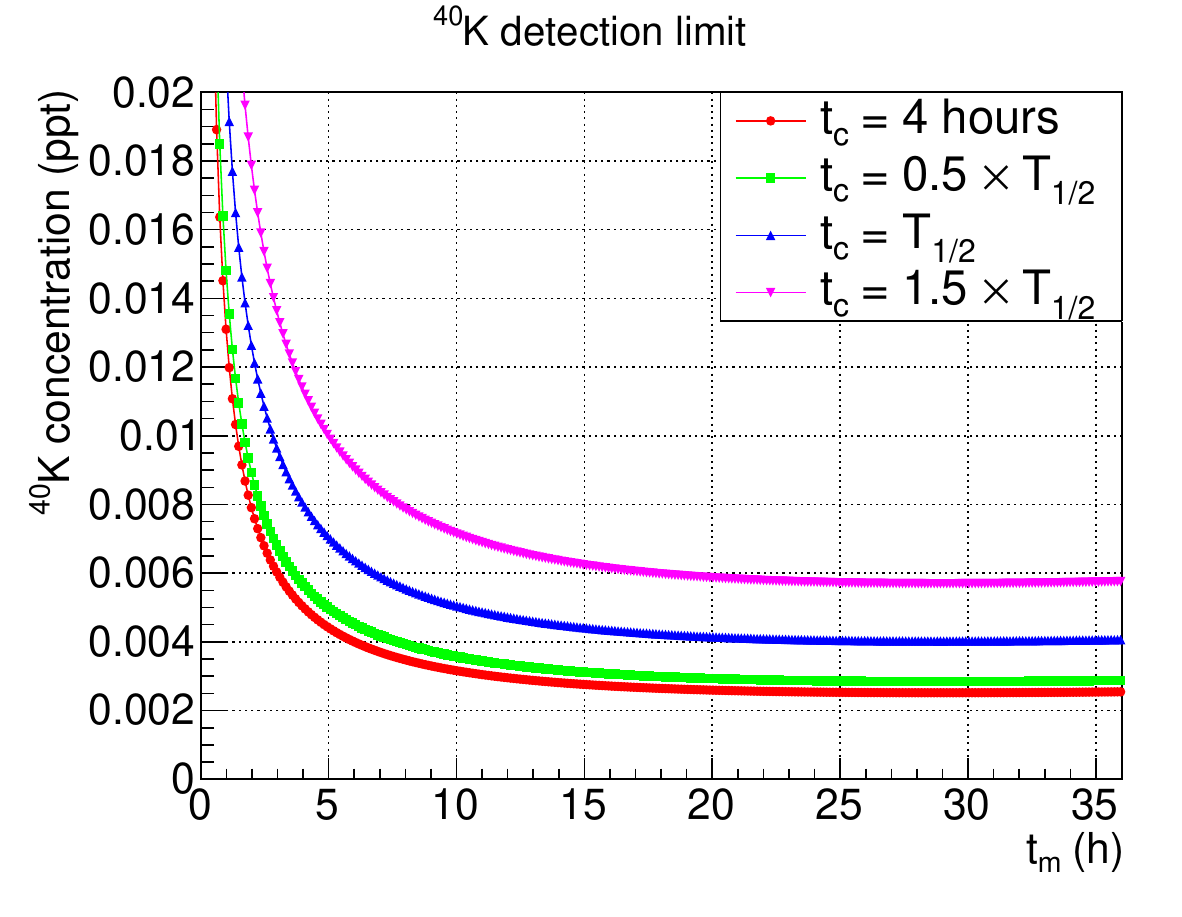}
\end{subfigure}
\hfill
\begin{subfigure}{0.45\textwidth}
    \includegraphics[width=\textwidth]{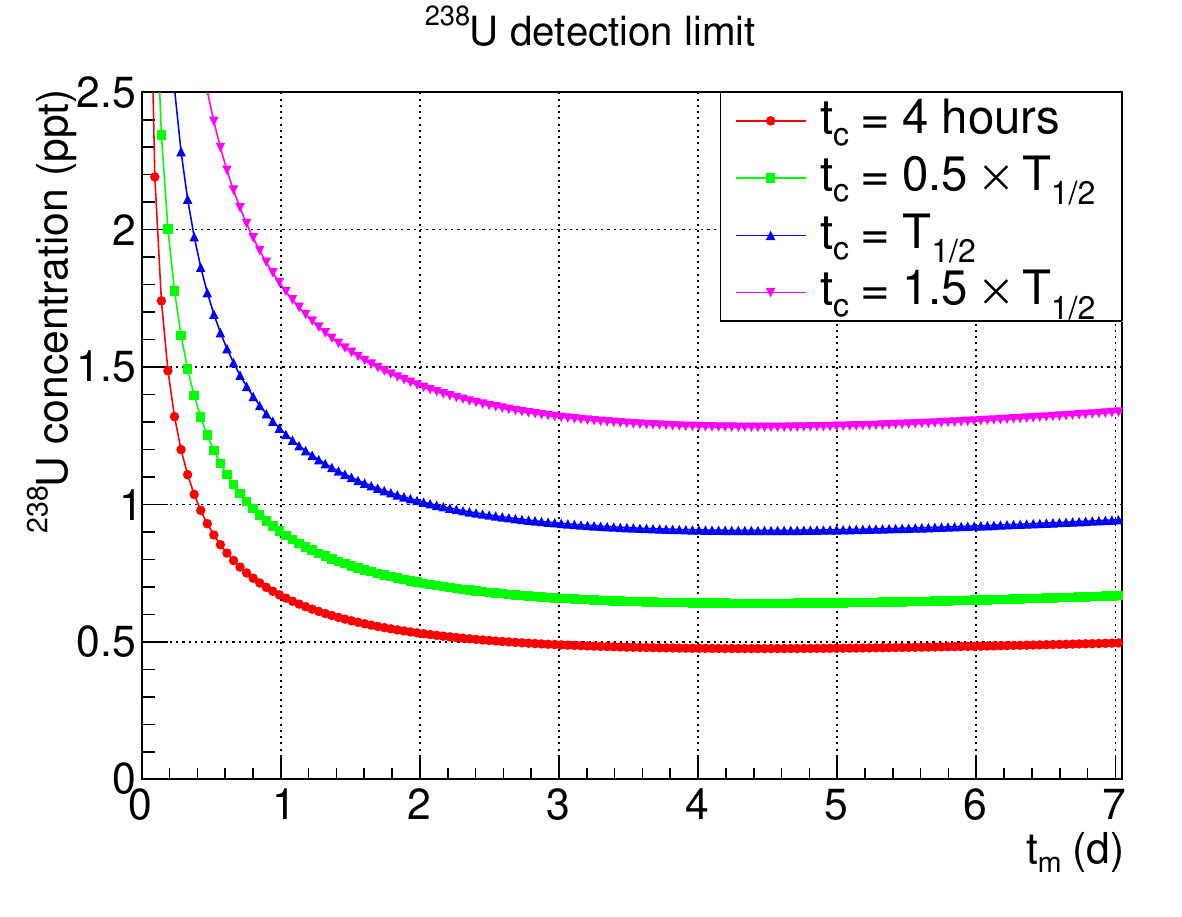}
\end{subfigure}
\hfill
\begin{subfigure}{0.45\textwidth}
    \includegraphics[width=\textwidth]{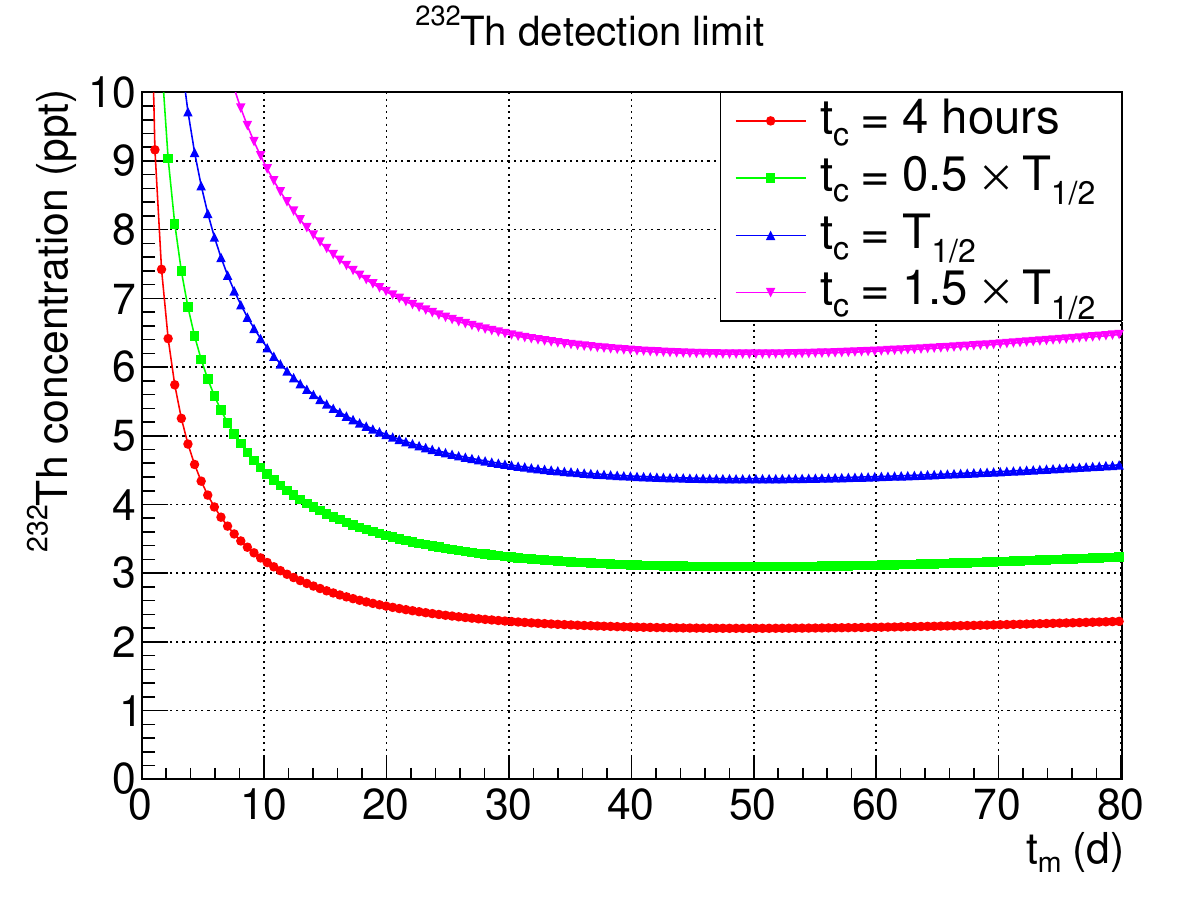}
\end{subfigure}
\caption{NAA detection limits achievable at the Radioactivity Laboratory of Milano-Bicocca (with the BeGE detector after 6 hours irradiation in the Lazy Susan facility of the TRIGA reactor of Pavia) for the concentration of the three natural contaminants in a generic plastic sample weighing 10\,g, in the absence of interfering activation products. For a detailed description of the plots refer to the text. The uncertainties on neutron flux, cross sections and the other parameters entering in Eq.~\ref{Eq:Sensitivity_1} are not represented here.}
\label{fig:Sensitivity}
\end{figure}

To calculate the number of decays $n_\text{dec}$ in Eq.~\ref{Eq:C_vs_nDec}, it is crucial to determine the detector efficiency at full energy for each $\gamma$-peak. This is achieved through a Monte Carlo (MC) simulation, based on GEANT4~\cite{Geant4}, that accurately models the HPGe detector, its surrounding setup, and the geometry of the measurement, including the shape and position of the sample. To validate the simulation, the calculated efficiencies were iteratively compared with experimental values obtained from certified multi-$\gamma$ reference sources. This process continued  until the agreement across the whole energy spectrum for all HPGe detectors in our laboratory was within 5\%. We conservatively regard this deviation as the systematic uncertainty associated with efficiency evaluation in our NAA campaigns.

\subsection{Expected NAA sensitivity at Milano-Bicocca}

Finally, we estimate the sensitivities for \k, \u, and \th, i.e. the smallest concentrations of natural contaminants that could potentially be detected with our experimental setup in the absence of interfering activation products. 
It is important to note that this estimate serves primarily for illustrative purposes and has a key limitation: it does not account for correlated background, which varies depending on the sample under investigation. 
For this calculation, we combine Eq.~\ref{Eq:R_vs_nDec}, \ref{Eq:Currie} and \ref{Eq:DetLimit} to determine the activation rate $R^*$ corresponding to the Currie's detection limit (see Sect.~\ref{sec:NAAsensitivity}), which ensures a clear identification of the $\gamma$-peak of the activation product. 
\begin{equation}
\label{Eq:Sensitivity_1}
R^* =  \dfrac{C^*}{\varepsilon_{\gamma} \, I_{\gamma} \, f(\Delta E)} \dfrac{\lambda}{(1-e^{-\lambda t_\text{irr}})e^{-\lambda t_\text{c}}(1-e^{-\lambda t_\text{m}})}
\end{equation}
Then, we use $R^*$ to calculate the corresponding number of target isotopes $\mathcal{N}^*=R^*/ (\sigma_\text{eff} \, \Phi)$  and, from these, the mass of the contaminant, which is finally divided by the mass of the sample to get the contaminant concentration.

In Fig.~\ref{fig:Sensitivity} we show the attainable sensitivity for the concentration of natural contaminants as a function of measuring time with the BEGe detector, considering a generic plastic sample with $m_{sample}=10$\,g and assuming a 6-hour irradiation in the Lazy Susan facility at the TRIGA reactor of Pavia. To this aim, we use the neutron flux and effective cross section values obtained from the measurements presented in Sect.~\ref{sec:systematics} (see Table~\ref{tab:Flux_1_2}, 2$^{nd}$ column, and Table~\ref{tab:XS_eff}).
For illustration purposes, we considered four different cooling times, ranging from a minimum of 4 hours to a maximum of 1.5 times the half-life of the activated product under study, and measuring times up to 3 times the same half-life. 

As one can see, concentrations ranging from a few ppq (for $^{40}$K) to a few ppt (for $^{232}$Th) are detectable with our experimental setup if the HPGe background is not increased by other activation products.

\section{Investigation of systematic uncertainties}
\label{sec:systematics}
Several factors contribute to the final uncertainty in the sensitivity of a given NAA campaign.
In this section, we present the results of a neutron activation experiment specifically carried out to investigate the systematic uncertainties due to self-shielding, moderation, and gradient effects in a typical irradiation configuration.

\subsection{Experiment description}
\label{sec:exp-description}

\begin{figure}[b]
\centering
\begin{subfigure}{0.23\textwidth}
    \includegraphics[width=\textwidth]{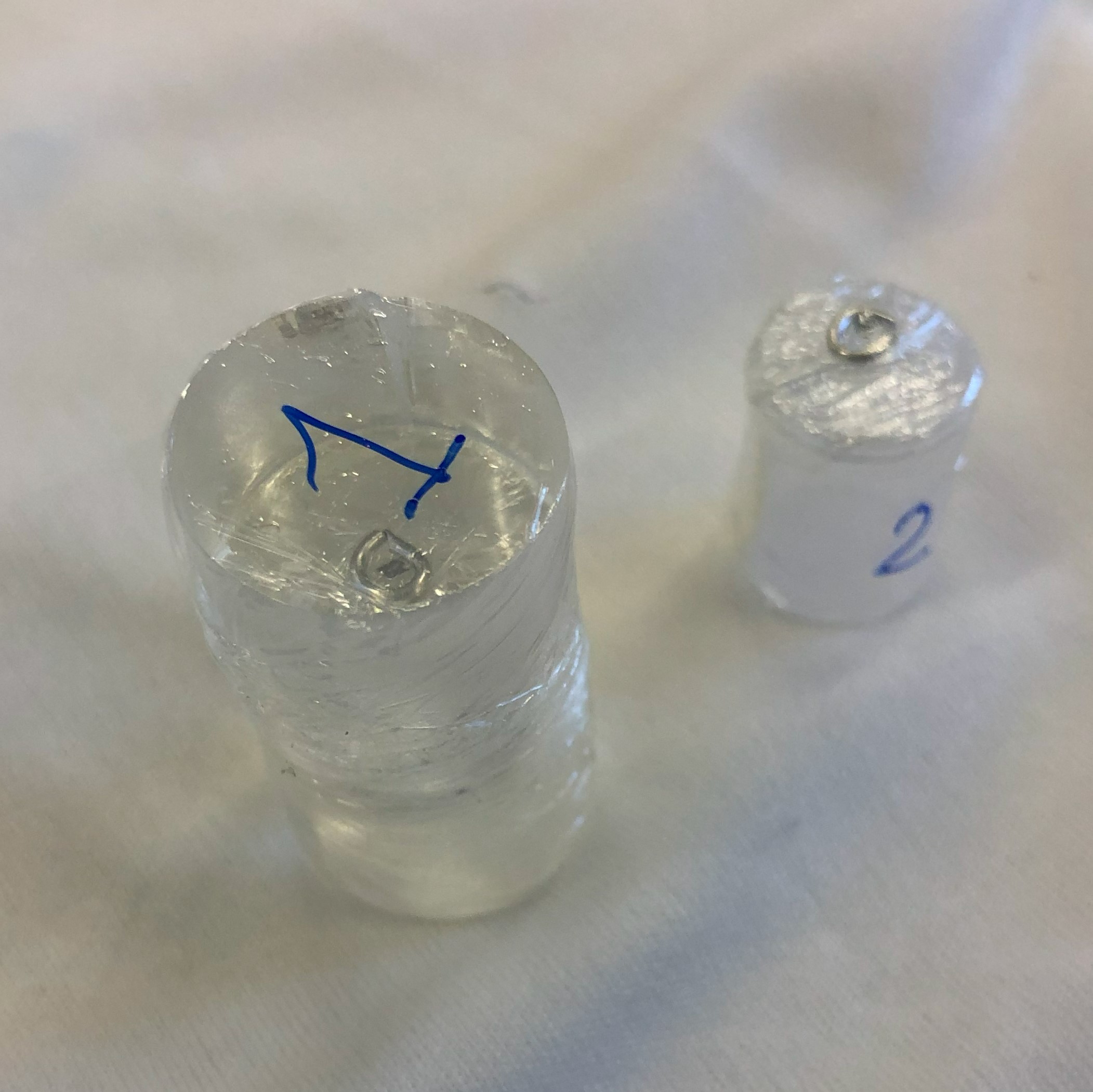}
    \caption{ }
    \label{fig:samples_a}
\end{subfigure}
~
\begin{subfigure}{0.23\textwidth}
    \includegraphics[width=\textwidth]{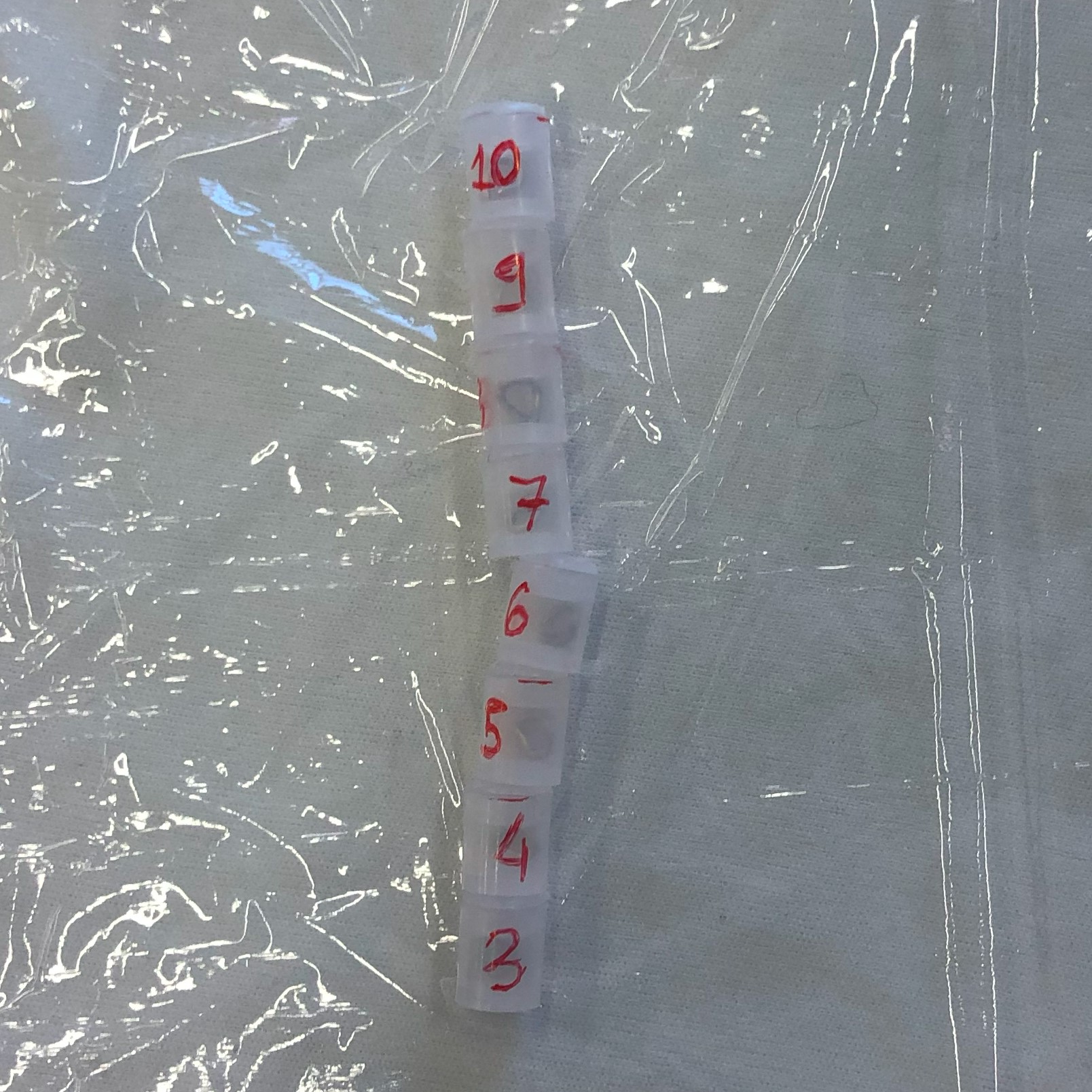}
    \caption{ }
    \label{fig:samples_b}
\end{subfigure}
~
\begin{subfigure}{0.23\textwidth}
    \includegraphics[width=\textwidth]{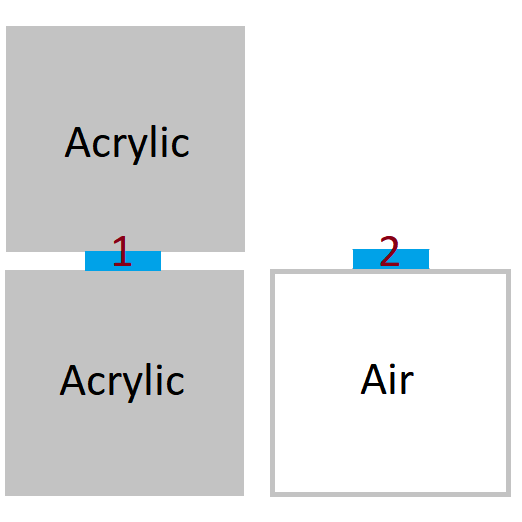}
    \caption{ }
    \label{fig:samples_c}
\end{subfigure}
~
\begin{subfigure}{0.23\textwidth}
    \includegraphics[width=\textwidth]{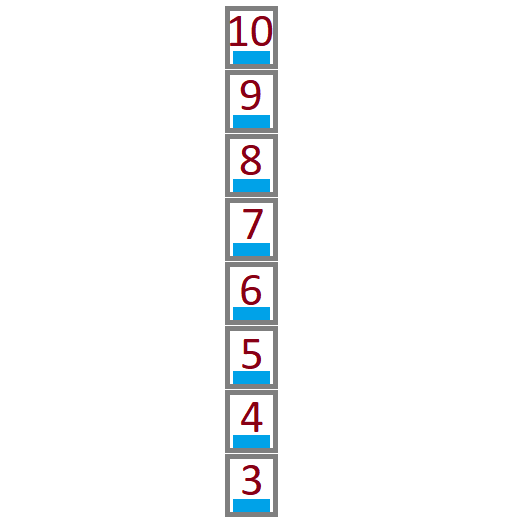}
    \caption{ }
    \label{fig:samples_d}
\end{subfigure}
\caption{(a) Al-Co and Al-Au samples sandwiched between two acrylic cylinders (label no.1) and positioned above an empty vial (label no.2). (b) Stack of 8 couples of Al-Co and Al-Au samples positioned inside small polyethylene vials. (c) Scheme (not to scale) of the setup in the top left photo. (d) Scheme (not to scale) of the setup in the top right photo.}
\label{fig:samples}
\end{figure}

\begin{table}[htb]
    \centering
    \renewcommand{\arraystretch}{1.2}
    \begin{tabular}{P{1cm}|P{1.5cm}|P{1.5cm}|P{1.4cm}|P{1cm}}
    Sample no.	&	Al-Co mass (mg)	&	Al-Au mass (mg)	&	 Irradiation Channel	&	Position (cm)	\\
    \hline
    1	&	20.97	&	0.585	&	LS--27	&	1.8	\\
    2	&	19.73	&	0.456	&	LS--28	&	1.8	\\
    3	&	19.39	&	0.422	&	LS--25	&	0.3	\\
    4	&	20.12	&	0.613	&	LS--25	&	1.3	\\
    5	&	19.66	&	0.398	&	LS--25	&	2.3	\\
    6	&	20.36	&	0.483	&	LS--25	&	3.4	\\
    7	&	20.68	&	0.523	&	LS--25	&	4.5	\\
    8	&	23.44	&	0.588	&	LS--25	&	5.5	\\
    9	&	18.73	&	0.386	&	LS--25	&	6.6	\\
    10	&	21.65	&	0.378	&	LS--25	&	7.7	\\
    \hline    
    \end{tabular}
    \caption{List of Al-Co (Co 0.1\% wt.) and Al-Au (Au 0.134\% wt.) samples irradiated in the Lazy Susan (LS) facility of the TRIGA Mark II reactor in Pavia. The experimental uncertainty of Al-Co mass is $\pm 0.01$~mg, whereas that of Al-Au samples is $\pm 0.003$~mg. In the last column we report the position of samples measured from the bottom of the holder with $\pm0.1$~cm precision.}
    \label{tab:masses}
\end{table}

We prepared samples made of certified Al-Co (Co 0.1\% wt.) and Al-Au (Au 0.134\% wt.) alloys, with the purpose of analyzing the following reactions:
\begin{align*}
    & ^{59}\text{Co} \xrightarrow{(n,\gamma)} \, ^{60}\text{Co} \xrightarrow[5.27\text{ yr}]{\beta^-} \, ^{60}\text{Ni} \\
    & ^{197}\text{Au} \xrightarrow{(n,\gamma)} \, ^{198}\text{Au} \xrightarrow[2.695\text{ d}]{\beta^-} \, ^{198}\text{Hg} \\
    & ^{27}\text{Al} \xrightarrow{(n,\alpha)} \, ^{24}\text{Na} \xrightarrow[14.96\text{ h}]{\beta^-}  \, ^{24}\text{Mg}
\end{align*}
These reactions allow to determine the three main components (thermal, intermediate, fast) of the neutron flux in a thermal reactor. Indeed, the $(n,\gamma)$ reactions on \cocn and \auuns, having different ratios between resonance integral and thermal cross section, allow to extract the thermal and intermediate flux components, whereas the $(n,\alpha)$ reaction on \alds, having a $\sim 5$~MeV energy threshold, can be used to measure the fast flux.

The irradiation was performed in July 2022 at the TRIGA Mark II reactor in Pavia and lasted for 6 hours at a nominal steady-state power of 250~kW. 
The first two couples of Al-Co and Al-Au samples (labelled as no.1 and no.2) were irradiated in two adjacent channels (LS--27 and LS--28, respectively) of the Lazy Susan facility.
As shown in Fig.~\ref{fig:samples_a}, the samples no.1 were sandwiched between two cylinders 1.8~cm high and 1.9~cm in diameter made of acrylic (C$_5$O$_2$H$_8$), whereas the samples no.2 were positioned at the same height as the first ones 
by means of an empty polyethylene vial. In this way, by comparing the activation rates of samples no.1 and no.2, we aim to investigate the effect of a moderator material surrounding the sample.
Moreover, in order to measure the vertical gradient of the activation rates, we irradiated a stack of 8 couples of Al-Co and Al-Au samples (no. from 3 to 10, Fig.~\ref{fig:samples_b}) in the channel LS--25.
In Table~\ref{tab:masses} we list the masses of the samples and we summarize their positions in the irradiation channels.

After the irradiation we measured all the activated samples with two HPGe detectors, named GeGEM and GeSilena\footnote{GeSilena is a p-type coaxial HPGe detector by Silena, with a relative efficiency of 30\% and an energy resolution of 1.9 keV at 1.33\,MeV.}, at our Radioactivity Laboratory. 
Performing double measurements of the same sample with two different detectors allows us to assess the accuracy of the efficiency evaluation in absolute activity measurements.

\subsection{Activation rate results}

\begin{figure}[b!]
\centering
\begin{subfigure}{0.48\textwidth}
    \includegraphics[width=\textwidth]{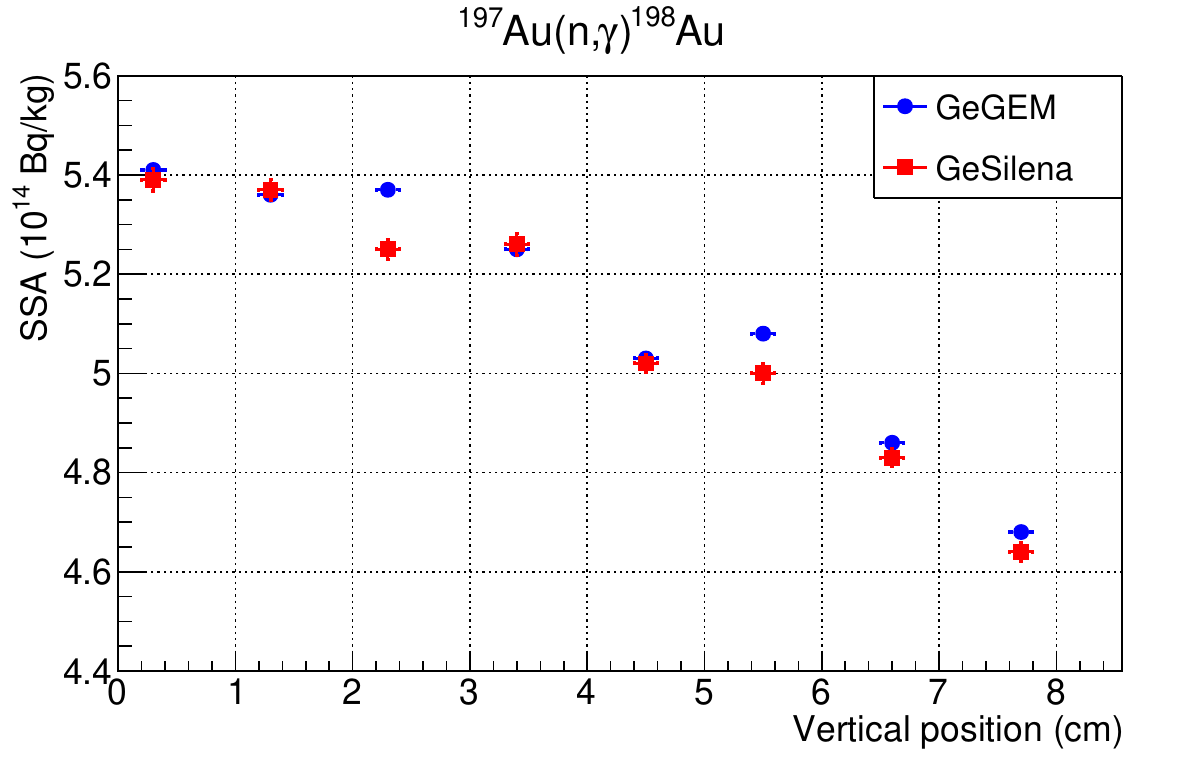}
\end{subfigure}
\hfill
\begin{subfigure}{0.48\textwidth}
    \includegraphics[width=\textwidth]{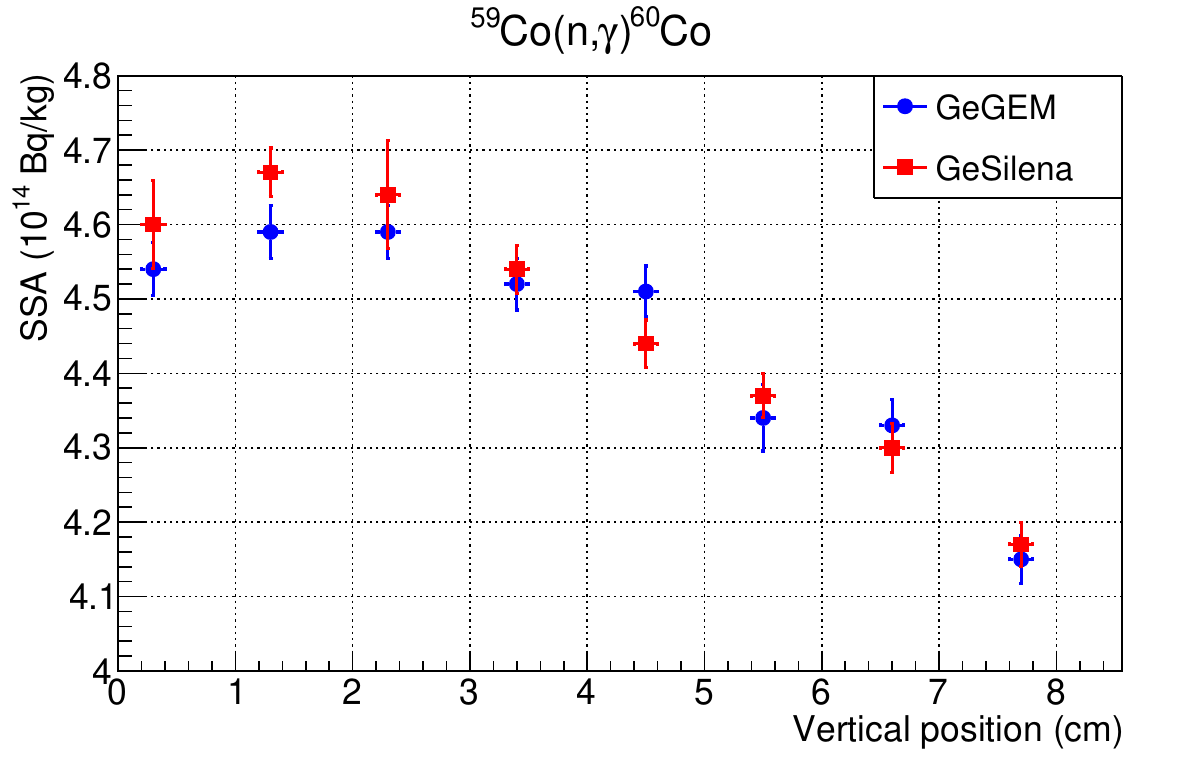}
\end{subfigure}
\begin{subfigure}{0.48\textwidth}
    \includegraphics[width=\textwidth]{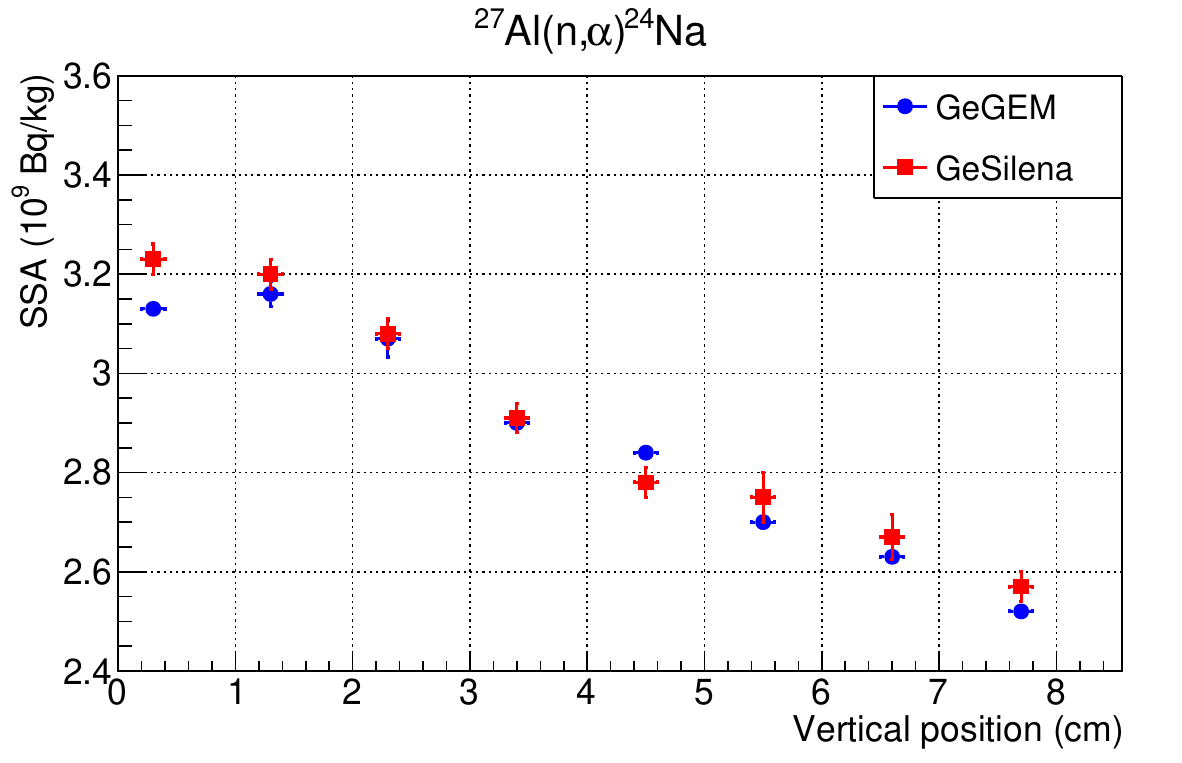}
\end{subfigure}
\caption{Vertical profiles of the Specific Saturation Activities of the radionuclides measured after irradiating Al-Au and Al-Co samples in the Lazy Susan facility (samples no.3 to no.10 in LS--25 channel). The results of $\gamma$-spectroscopy measurements performed with two HPGe detectors (GeGEM and GeSilena) are shown with different markers and colors.}
\label{fig:SSA}
\end{figure}

\begin{table*}[htb!]
    \centering
    \renewcommand{\arraystretch}{1.2}
    \begin{tabular}{c|c|c|c}
    Reaction             &	SSA (Bq/kg) \textit{w/ moderator}     &	SSA	(Bq/kg)	\textit{w/o moderator}  &	 $\Delta$(\%)	\\
    \hline
    \auuns\ngamma\auuno	 &  $(5.679\pm0.009)\times10^{14}$ & $(5.167\pm0.009)\times10^{14}$ &    $9.9\pm0.2$           \\
    \cocn\ngamma\cosz    &  $(5.018\pm0.029)\times10^{14}$ & $(4.493\pm0.024)\times10^{14}$ &    $11.7\pm0.8$           \\
    \alds\nalfa\nadq     &  $(2.973\pm0.011)\times10^{9}$  & $(2.889\pm0.011)\times10^{9}$  &    $2.9\pm0.5$           \\
    \hline    
    \end{tabular}
    \caption{Specific Saturation Activities of the three reactions measured in the Al-Au and Al-Co samples no.1 (positioned between two acrylic moderators) and no.2 (without moderators) irradiated in the Lazy Susan facility. In the last column we report the relative difference of SSA in sample no.1 with respect to sample no.2.}
    \label{tab:SSA_1_2}
\end{table*}

\begin{figure}[htb]
\centering
\includegraphics[width=0.48\textwidth]{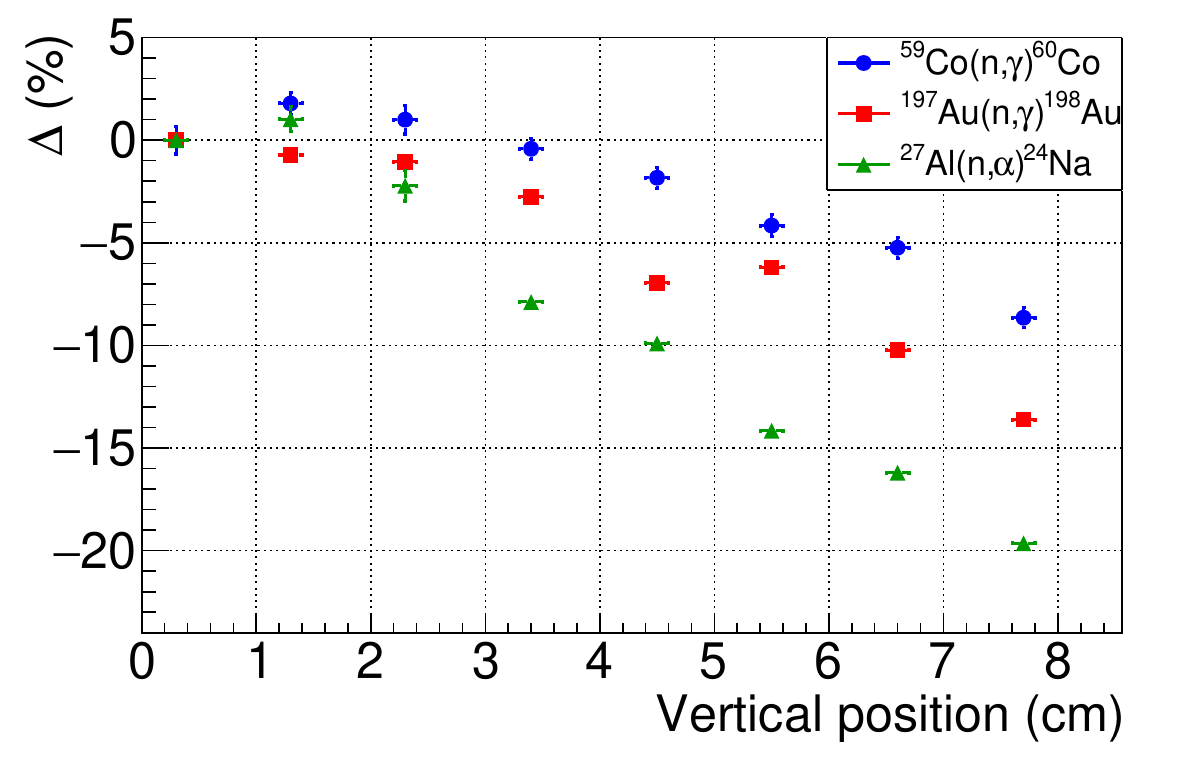}
\caption{Relative differences of the Specific Saturation Activities of \auuno, \cosz, and \nadq, calculated with respect to the ones measured in lowest irradiation position of the LS--25 channel.}
\label{fig:RelDiff}
\end{figure}

We exploit the analysis methods described in Sect.\ref{sec:data_analysis} to get the number of counts $C$ in the $\gamma$-peaks of HPGe spectra and to evaluate the corresponding detection efficiencies $\varepsilon_{\gamma}$ through MC simulations. 
Particularly, we analyze the $\gamma$-lines at 1173~keV and 1333~keV emitted by \cosz, the one at 412~keV of \auuno, and those at 1369~keV and 2754~keV of \nadq.
By combining Eq.~\ref{Eq:R_vs_nDec} and ~\ref{Eq:C_vs_nDec} we get the activation rates $R$ for the three reactions of interest. 

First of all, we compare the results obtained for $R$ from the measurements on the two different HPGe detectors.
The differences are within $\pm3\%$ for all samples, pointing out that our MC tool for $\varepsilon_{\gamma}$ estimation can reach the same accuracy level certified for the activity of the multi-$\gamma$ sources used in our laboratory for HPGe calibration. 

Then, in order to compare the activation results obtained from the different samples, we divide $R$ by the mass $m$ of the target element in each sample. The activation rate per unit mass $R/m$ is usually referred to as Specific Saturation Activity (SSA), where \textit{specific} means ``per unit mass of the target element" and \textit{saturation} refers to the condition (reached through an irradiation much longer than the half-life of the activation product) in which the decay rate equals the activation rate.

In Table~\ref{tab:SSA_1_2}, we report our best estimates\footnote{Computed as the weighted average of the $R/m$ results from the two HPGe measurements.} for the SSA of samples no.1 and no.2 irradiated with and without a surrounding moderator, respectively. In the last column, we highlight that the \ngamma reactions are characterized by activation rates $\sim10\%$ higher when the samples are surrounded by a moderator, whereas the difference is lower for the \nalfa reaction induced by fast neutrons.
Since the samples no.1 and no.2 have been irradiated in two adjacent channels of the Lazy Susan facility, we have considered the hypothesis that there is non-uniformity of flux between the channels. However, given the simulation results shown later in Section~\ref{sec:syst_discussion}, we can expect flux variations of $\mathcal{O}(1\%)$ between channels LS--27 and LS--28.
Therefore, we can state the observed $\sim10\%$ differences for \ngamma activation rates are due to the effect of the moderator. In particular, they can be interpreted as the consequence of a local flux enhancement due to elastic scattering in the moderator, which causes more neutrons, especially thermal ones, to pass back and forth through the sample.

In Fig.~\ref{fig:SSA}, we show the vertical profiles of SSA for the three measured reactions, as they result from the measurements of samples from no.3 to no.10. From the analysis of these profiles it is possible to conclude that the activation rates are compatible with each other in the lower 2 or 3~cm, whereas they exhibit a clear decreasing trend moving to the top.
Fig.~\ref{fig:RelDiff} shows the relative differences of the SSA at each position compared to the SSA measured in the lowest position of the LS-25 channel.

\subsection{Neutron flux results}

\begin{figure*}[htb]
\centering
\begin{subfigure}{0.48\textwidth}
    \includegraphics[width=\textwidth]{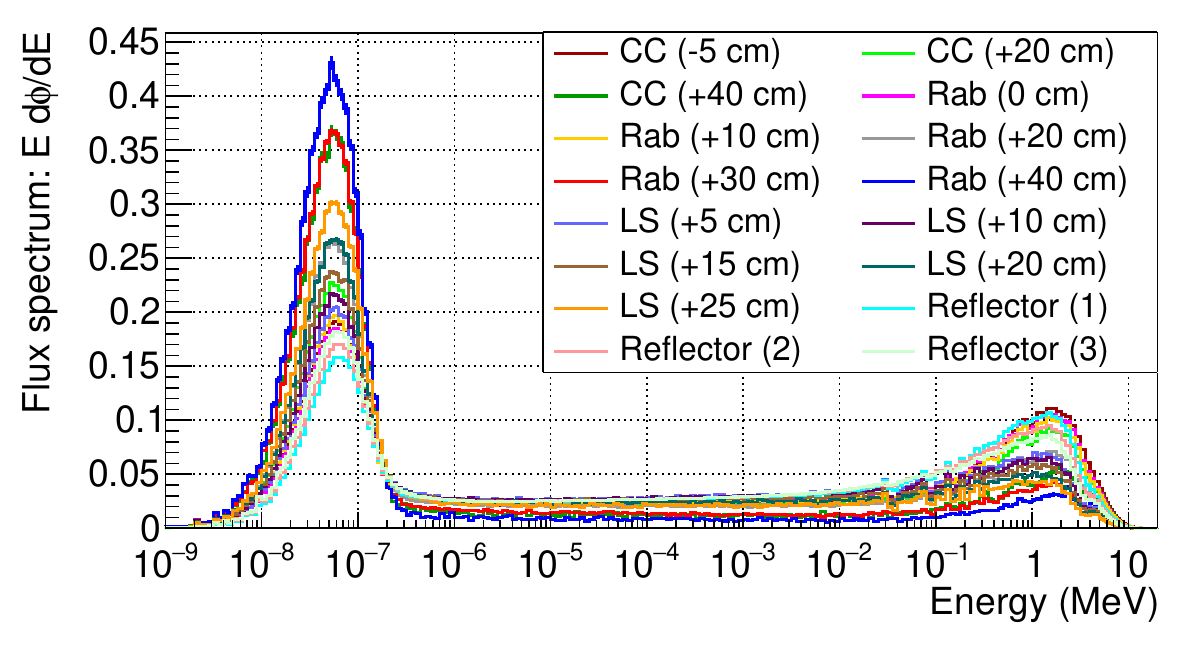}
    \caption{ }
    \label{fig:GuessSpectra}
\end{subfigure}
\hfill
\begin{subfigure}{0.48\textwidth}
    \includegraphics[width=\textwidth]{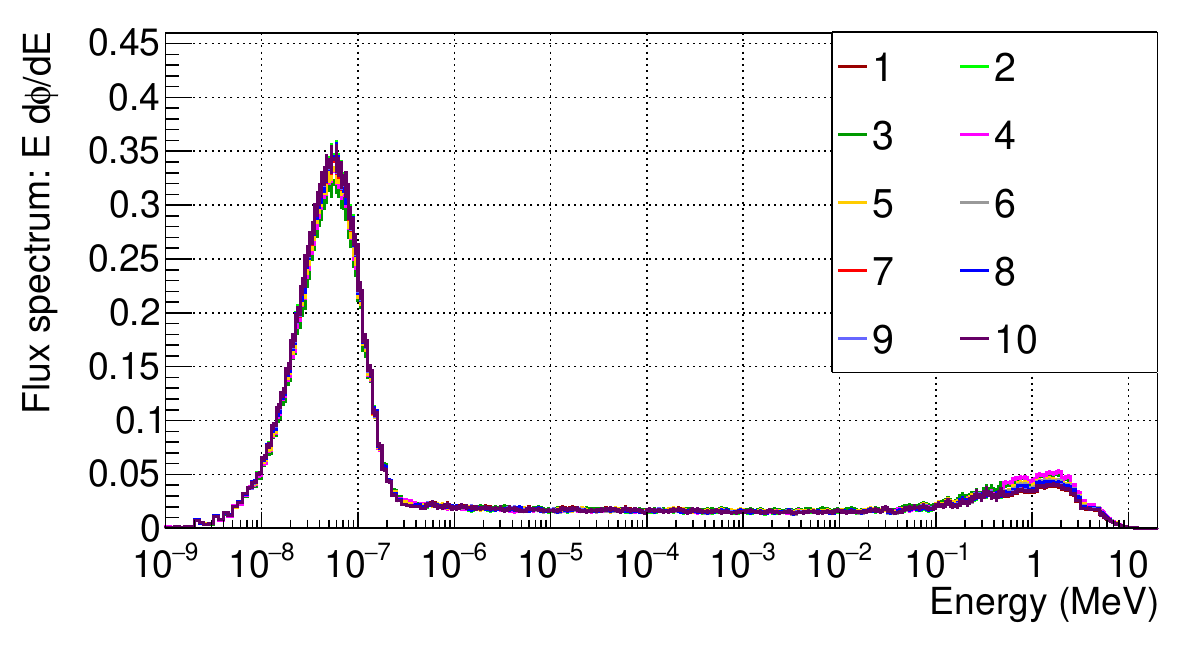}
    \caption{ }
    \label{fig:UnfoldedSpectra}
\end{subfigure}
\caption{(a) Neutron flux spectra, shown in lethargic scale and normalized to 1, obtained from the MCNP simulation model of the TRIGA Mark II reactor in Pavia at various distances from the center of the core along the z-axis in the Central Channel (CC), Rabbit (Rab), and Lazy Susan (LS) irradiation facilities, and in 3 annulus volumes in the reflector region between the core and the Lazy Susan (for a description of the TRIGA reactor geometry refer to~\cite{AbsoluteFlux}). All the other spectra (not shown here) belonging to the set of \textit{guess} spectra used in the unfolding analysis are characterized by similar shapes, comprised between the most and least thermalized spectra shown here. (b) Spectra, normalized to 1, obtained at the end of the unfolding analysis of the 10 samples irradiated in different positions and conditions in the LS facility described in this paper.}
\end{figure*}

The next step in the analysis consists on the characterization of the neutron flux and its spectrum in the different irradiation configurations and positions.
In this way, we get the information needed to assess the systematic uncertainties affecting the activation measurements of any nuclide in the Lazy Susan facility.

For this analysis, we divide the neutron flux spectrum into three energy groups:
\begin{enumerate}
    \item $(0 - 0.5~\text{eV})$ for thermal neutrons;
    \item $(0.5~\text{eV} - 0.5~\text{MeV})$ for epithermal neutrons;
    \item $(0.5~\text{MeV} - 20~\text{MeV})$ for fast neutrons. 
\end{enumerate}

To determine the fluxes for each group, we use the Bayesian-unfolding Toolkit for Multi-foil Activation with Neutrons (BaTMAN)~\cite{BaTMAN}, originally developed to analyze a set of activation data collected at the TRIGA reactor in Pavia in 2012~\cite{AbsoluteFlux,BayesianSpectrum}, and subsequently applied to characterize the neutron fluxes at the TRIGA reactor in Rome~\cite{Chiesa:2019buj} and at spallation source beamlines~\cite{Chiesa:2018ane,Chiesa:2021ven}.

The BaTMAN unfolding algorithm solves the following system of linear equations -- one for each activation reaction $j$ -- obtained by rewriting Eq.~\ref{Eq:ActRate} as:
\begin{equation}
\label{Eq:ActRateDiscrete}
\dfrac{R_j}{\mathcal{N}_j} =  \sum_{i=1}^{3} \sigma_{ij} \phi_i 
\end{equation}
where $\sigma_{ij}$ are the effective cross sections (Eq.~\ref{Eq:XSeff}) calculated in each energy group, and $\phi_i$ are the three unknown variables of the system, corresponding to the thermal, epithermal and fast fluxes.
The $\sigma_{ij}$ coefficients must be calculated assuming a \textit{guess} spectral shape $\varphi(E)$ in the energy range of each group. 
For this purpose, we exploit the MCNP~\cite{mcnp} simulation model of the TRIGA Mark II reactor in Pavia developed and validated a few years ago~\cite{FirstReactorMODEL,FreddoPulito,CaldoPulito,Burnup,FluxDistribution}.
Particularly, we collected a set of different simulated neutron spectra by tallying the neutron flux in various positions of the reactor core, reflector and irradiation facilities. 
As shown in Fig.~\ref{fig:GuessSpectra}, all these spectra are characterized by common features, because in thermal reactors the fast neutrons -- emitted by fission reactions with the so-called Watt spectrum~\cite{watt} -- undergo a moderation process that generates a flux $\varphi(E) \propto E^{-\beta}$ (with $\beta \simeq 1$) in the intermediate range and a Maxwell distribution at thermal energies~\cite{lamarsh}. 
It is worth noting that even if the ratios between thermal, epithermal and fast flux integrals can vary significantly, the \textit{intra-group} spectral shapes are similar. Thus, as a first approximation, we can calculate $\sigma_{ij}$ values by using as \textit{guess} any of the spectra belonging to the aforementioned set.

The BaTMAN unfolding algorithm solves the system in Eq.~\ref{Eq:ActRateDiscrete} with a Bayesian statistical approach.
Compared to other unfolding techniques~\cite{Reginatto2010}, this method allows to rigorously propagate the experimental uncertainties, to select the physical solutions of the problem using the Priors and to analyze the correlations between the resulting $\phi_i$.
For each irradiation position, we provide as input:
\begin{itemize}
    \item the SSA data of the three activation reactions;
    \item the SSA uncertainty, calculated as the quadrature sum of the statistical uncertainty plus a 3\% systematic uncertainty on detector efficiency;
    \item the $\sigma(E)$ cross sections, taken from the ENDF/B-VII data libraries~\cite{ENDF}, with associated uncertainties from the BNL-98403-2012-JA report~\cite{PRITYCHENKO20123120};
    \item a \textit{guess} spectrum randomly selected from the set of spectra produced with MCNP;
\end{itemize}
The multi-dimensional Posterior probability density function $p(\phi_{i}|R_j,\sigma_{ij})$, given by the product of Gaussian Likelihoods defined from the experimental data and uniform Priors in the positive range for the $\phi_{i}$ unknowns, is then sampled through the Just Another Gibbs Sampler (JAGS) tool~\cite{JAGS,JAGS_manual}, which exploits the Markov Chains Monte Carlo (MCMC) method~\cite{Gelman}. 
At the end, through the marginalization of the multi-dimensional Posterior, we get the probability density function of each flux group $\phi_i$ and the correlations among them. The means and standard deviations of the marginalized Posteriors are finally computed to provide a three-group neutron flux unfolding result.
With this unfolding method, even if the results have some degree of dependence on the \textit{intra-group} spectral shape used for $\sigma_{ij}$ calculation, there is no constraint on $\phi_i$ variables, that are left free to converge on any flux value in the positive range.

In order to identify, among spectra produced with the MCNP simulations, the one that better fits the unfolding results, we implement a $\chi^2$ minimization and we apply the iterative procedure described in Ref.~\cite{Chiesa:2019buj}. 
After a few iterations ($<10$), the unfolding solution stabilizes at values consistent with one of the spectra available in the MCNP set.

\begin{table}[b]
    \centering
    \renewcommand{\arraystretch}{1.2}
    \begin{tabular}{c|c|c}
    Energy              &	Flux \textit{w/ moderator}     &	Flux \textit{w/o moderator}   \\
    group             &	 ($10^{12}\,\text{cm}^{-2}\text{s}^{-1}$)      &	($10^{12}\,\text{cm}^{-2}\text{s}^{-1}$)   \\
    \hline
    Thermal	 &  $1.52\pm0.06$ & $1.37\pm0.05$            \\
    Epithermal    &  $0.58\pm0.07$ & $0.49\pm0.06$ \\
    Fast     &  $0.171\pm0.008$  & $0.176\pm0.008$  \\
    \hline    
    Total     &  $2.26\pm0.05$  & $2.03\pm0.04$  \\
    \hline
    \end{tabular}
    \caption{Neutron flux results obtained from the unfolding of the activation measurements of samples no.1 (positioned between two acrylic moderators) and no.2 (without moderators) irradiated in the Lazy Susan facility.}
    \label{tab:Flux_1_2}
\end{table}

\begin{figure}[htb]
\centering
\includegraphics[width=0.48\textwidth]{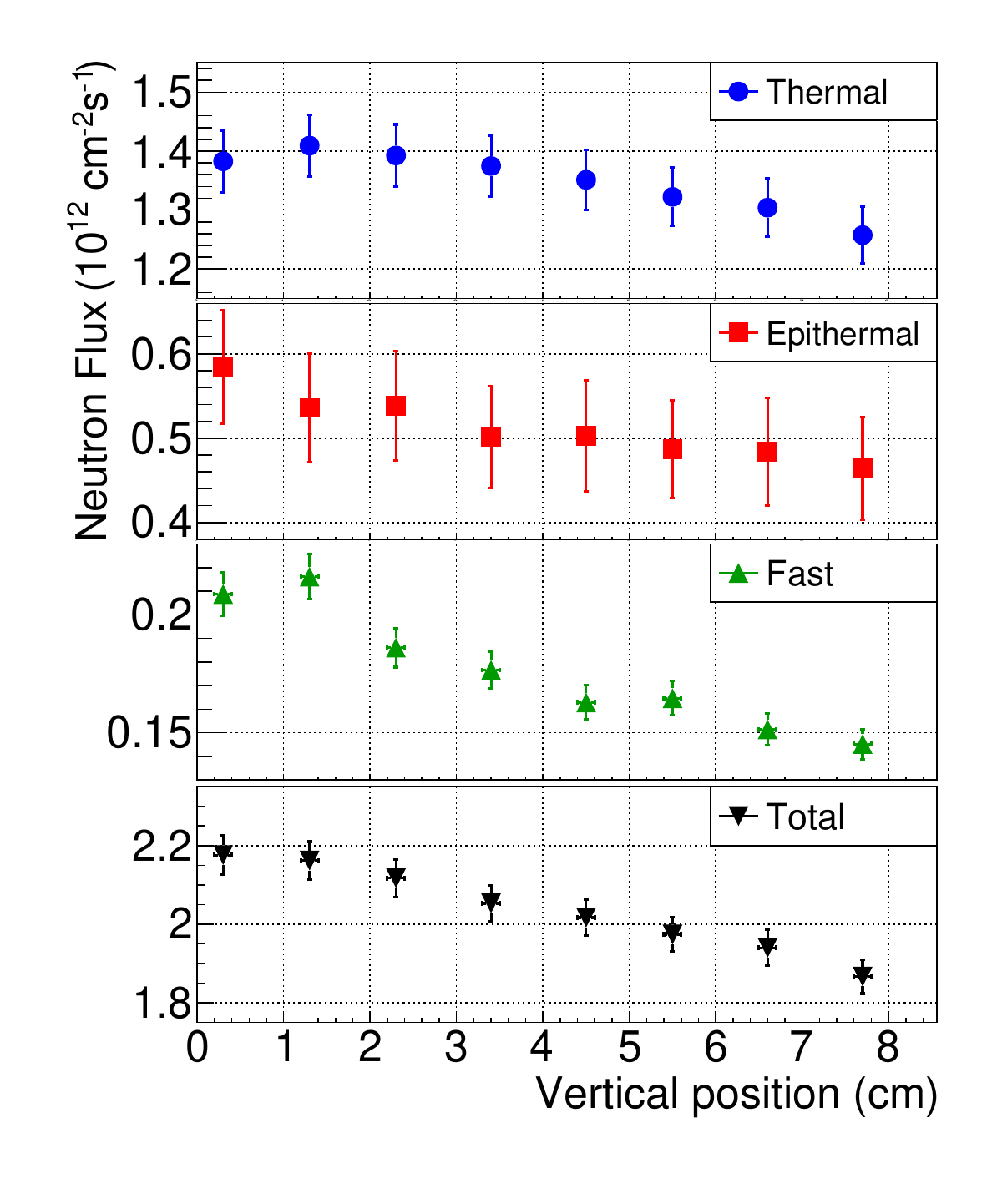}
\caption{Neutron flux unfolding results obtained from the activation measurements of the vertical stack of samples in the LS--25 channel.}
\label{fig:FluxProfile}
\end{figure}

To graphically represent the unfolded spectrum, we normalize the intra-group spectral shapes of the last guess spectrum to the thermal, epithermal and fast flux values obtained after running JAGS.
In Fig.~\ref{fig:UnfoldedSpectra}, we show the unfolded spectra --normalized to 1-- for all the irradiation positions. 
Two aspects are worth noting.
First, the spectra do not exhibit discontinuities at the group boundaries even if no continuity condition is imposed in the unfolding algorithm, therefore this can be considered as an indication of the robustness and self-consistency of the results.
Second, all the unfolded spectra are very similar to each other, both when comparing the results from the irradiation with and without the moderator (1 vs 2), and when analyzing the vertical profile (from 3 to 10).

Conversely, we find appreciable differences when the neutron flux results are considered on an absolute scale. 
Indeed, as shown in Table~\ref{tab:Flux_1_2}, a relative difference of $(12\pm3)\%$ is obtained for the total flux measured with or without the moderator. This difference arises from a variation of the thermal and epithermal flux, whereas the fast flux is compatible in the two measurements, as expected.
We point out that the thermal and epithermal fluxes are highly anti-correlated (the correlation coefficient is $-0.7$), so their sum can be determined with a smaller uncertainty ($\sim2.5\%$) than each of them taken separately.

In Fig.~\ref{fig:FluxProfile}, we show the flux profile as a function of the vertical position in the Lazy Susan irradiation channel. A clear decreasing trend characterizes all three flux groups. The relative differences between the fluxes in the higher versus the lower position are: $(-9\pm3)\%$, $(-21\pm10)\%$, and  $(-30\pm3)\%$ for the thermal, epithermal and fast component, respectively. 
The relatively large uncertainty affecting the epithermal group is mainly due to the strong anti-correlation (always $\sim-0.7$) with the thermal one.
The maximum variation of the total flux, dominated by the thermal and epithermal components which together constitute over 90\% of it, is $(-14\pm2)\%$.

If we compare the results obtained here with those published in Ref.~\cite{DiLuziok0PV:2019}, in which the parameters used in the framework of the $k_0$-method have been measured along the vertical axis of the irradiation facilities of the TRIGA reactor in Pavia, we find that the gradients of the thermal and epithermal fluxes are in good agreement.
Given the different formalism used here with respect to that introduced in the framework of the $k_0$-method, a direct comparison on an absolute scale of the thermal and epithermal fluxes taken individually is not straightforward. However their sum can be easily compared using the following equation:
\begin{equation}
    \phi_1+\phi_2 = \Phi_{th} + \Phi_e \int_{0.5 \text{eV}}^{0.5 \text{MeV}} \dfrac{1}{E^{1+\alpha}} dE 
    \label{Eq:k0_cfr}
\end{equation}
where $\phi_1$ and $\phi_2$ are the thermal and epithermal fluxes measured in this work, whereas $\Phi_{th}$, $\Phi_e$ and $\alpha$ are the flux parameters determined in Ref.~\cite{DiLuziok0PV:2019} for the $k_0$-method application.
The comparison of the results performed through Eq.~\ref{Eq:k0_cfr} gives us differences below the 2.5\% statistical uncertainty affecting $\phi_1+\phi_2$, highlighting that the technique that we used to characterize and unfold the neutron fluxes is accurate and reliable.

\subsection{Discussion}
\label{sec:syst_discussion}

A first point worthy of discussion is that the neutron fluxes obtained from the unfolding analysis on the one hand are compatible with those published in Ref.~\cite{DiLuziok0PV:2019}, but on the other they exhibit non-negligible differences compared to the predictions of the MCNP model of the TRIGA Mark II reactor in Pavia.
In particular, the Monte Carlo model predicts a total neutron flux of $(2.76 \pm 0.04)\times 10^{12}$~cm$^{-2}$s$^{-1}$ at the bottom of the Lazy Susan facility~\cite{AbsoluteFlux}, thus being $\sim30\%$ higher than the flux measured in this work without moderator. 
Furthermore, the MCNP simulated spectrum presents a relative proportion between thermal, epithermal and fast fluxes of $43\%:45\%:12\%$, which is not compatible with our unfolding analysis which gives us, in all irradiation positions and configurations, group fractions falling into the following ranges (in brackets we report the statistical uncertainty):
\begin{itemize}
    \item thermal: $64\% - 67\%$ ($\pm3\%$)
    \item epithermal: $24\% - 27\%$ ($\pm3\%$)
    \item fast: $7.6\% -10\%$ ($\pm0.4\%$)
\end{itemize}
This discrepancy, perhaps due to too approximate modeling of the material of which the reflector is made, had already been noticed in the past and its investigation is beyond the scope of this work.
Indeed, this is the main reason that convinced us to carry out experimental measurements to investigate the systematic uncertainties, rather than relying solely on the Monte Carlo model of the reactor.
However, despite these issues, the MCNP model has been an extremely valuable tool for conducting the unfolding analysis and to get the neutron flux spectra shown in Fig.~\ref{fig:UnfoldedSpectra}.

\begin{table*}[htb!]
    \centering
    \renewcommand{\arraystretch}{1.2}
    \begin{tabular}{c|c|c|c|c}
    Nuclide & $\sigma_\text{th}$ (barn) & $\sigma_\text{epi}$ (barn) & $\sigma_\text{fast}$ (barn) & $\sigma_\text{eff}$ (barn) \\
    \hline
    \kqu & $1.19\pm0.02$ & $0.067\pm0.004$ & $0.0022\pm0.0001$ & $0.81\pm0.02$\\
    \th  & $5.94\pm0.02$ & $5.57\pm0.14$ & $0.096\pm0.002$ & $5.34\pm0.17$\\
    \u   & $2.20\pm0.02$ & $18.5\pm0.5$    & $0.071\pm0.001$ & $6.09\pm0.45$\\
    \hline
    \end{tabular}
    \caption{Effective cross sections obtained from the unfolded spectra of the Lazy Susan facility. All the values obtained from the 10 different spectra shown in Fig.~\ref{fig:UnfoldedSpectra} fall within the quoted uncertainty ranges.    
    The \textit{group} effective cross sections calculated in the thermal, epithermal and fast range, are labeled as $\sigma_\text{th}$, $\sigma_\text{epi}$, and $\sigma_\text{fast}$. Their uncertainties include that of the cross section data taken from~\cite{PRITYCHENKO20123120} and that of the \textit{guess} MCNP spectra. 
    In the last column we report the effective cross sections calculated from 0 to 20 MeV. Their uncertainties are obtained by combining that of neutron flux groups (taking into account their correlations) with that of cross section data.}
    \label{tab:XS_eff}
\end{table*}

A direct application of the neutron flux characterization conducted with this methodology, that combines the accuracy of the experimental measurements with the precision of the Monte Carlo simulations in determining the intra-group spectral shapes, is the calculation of the effective cross sections and the activation rates of any nuclide of interest.
We present an example of such application for the most relevant nuclides in the framework of radiopurity assays measurements.
We use all the unfolded spectra reported in Fig.~\ref{fig:UnfoldedSpectra} to calculate the effective cross section for (n,$\gamma$) reactions on \kqu, \th and \u.
By calculating the effective cross sections with the different spectra, it turns out that they are compatible with each other within their uncertainties. This outcome is easily understandable given the similarity of the unfolded spectra. In Table~\ref{tab:XS_eff} we report the average values of the effective cross sections.

To compare the $(n,\gamma)$ reaction rates on \kqu, \th, and \u targets when the neutron irradiation is performed with or without a $\sim2$~cm thick moderator material, we combine the group effective cross sections with the unfolding results in Table~\ref{tab:Flux_1_2}. 
As expected, the activation rates are higher in presence of a moderator. In the second column of Table~\ref{tab:Bias_moderator} we show the ratios between the activation rates of \kqu, \th and \u with/without moderator.
These results highlight that a not negligible systematic bias can affect NAA results when the activated sample is made of a moderator material (with thickness of the order of 2~cm) whereas the co-irradiated \textit{standard} is a smaller sample (size of few mm). 
We point out that the bias can be different depending on the fraction with which the activation reactions are induced by thermal and epithermal neutrons, respectively. 
These activation fractions may vary considerably from reaction to reaction. In the right columns of  Table~\ref{tab:Bias_moderator} we list the thermal and epithermal activation fractions for the three reactions under consideration. 
It is interesting to note that \u (differently from \kqu and \th) is activated mainly by epithermal neutrons, whose flux was determined with less precision. This explains why the bias affecting \u activation was determined with greater uncertainty with respect to \kqu and \th ones.

\begin{table}[htb]
    \centering
    \renewcommand{\arraystretch}{1.2}
    \begin{tabular}{c|c|c|c}
    \multirow{2}{*}{Nuclide} & \multirow{2}{*}{$\dfrac{R_\text{w/ moderator}}{R_\text{w/o moderator}}$} &  Thermal & Epithermal\\
    & & activation & activation \\
    \hline
    \kqu & $1.11\pm0.06$ & 98\% & 2\% \\
    \th   & $1.13\pm0.04$ & 74\% & 26\% \\
    \u  & $1.17\pm0.14$ & 24\% & 76\% \\
    \hline
    \end{tabular}
    \caption{($2^{nd}$ \textit{column}) Estimate of the activation rate ratio when neutron irradiation is performed with or without a surrounding moderator. The uncertainties have been calculated by propagating the neutron flux uncertainties only, because the cross section ones almost completely cancel out when calculating the ratio between activation rates. ($3^{rd}$ and $4^{th}$ \textit{columns}) Fractions with which the activation reactions are induced by thermal and epithermal neutrons in the Lazy Susan facility of the TRIGA Mark II reactor in Pavia.}
    \label{tab:Bias_moderator}
\end{table}

\begin{figure}[htb]
\centering
\includegraphics[width=0.48\textwidth]{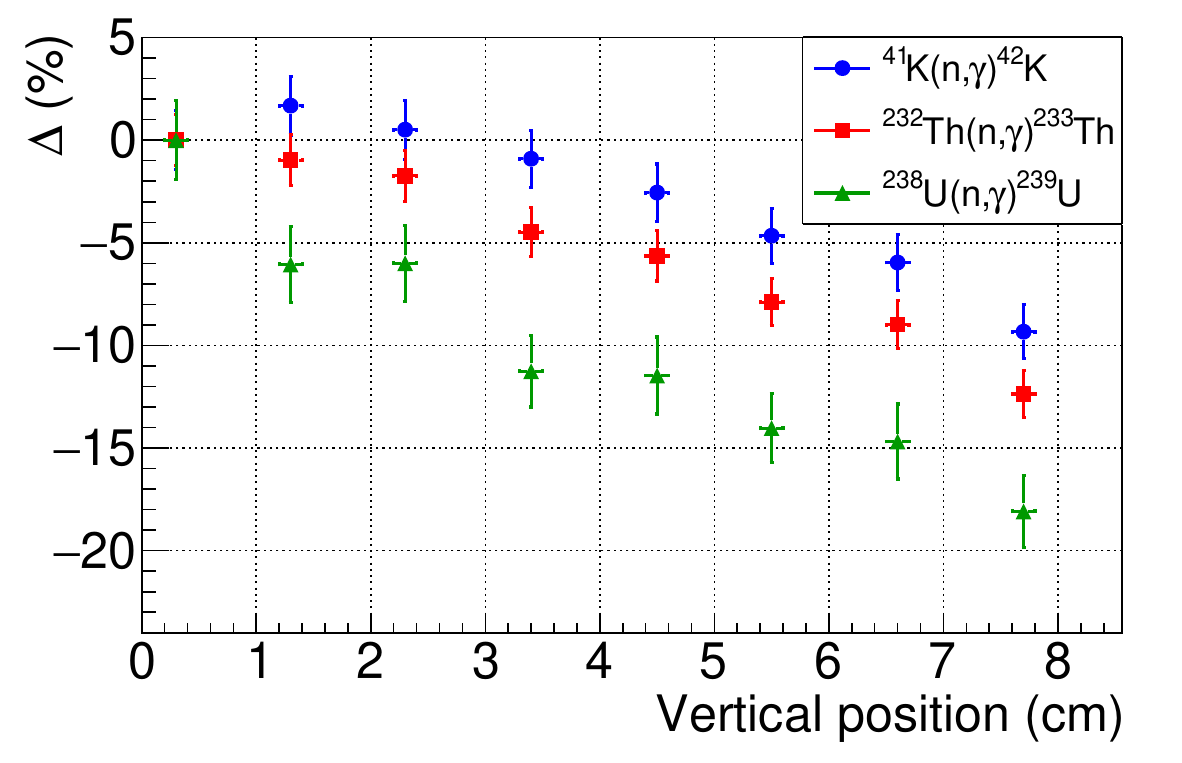}
\caption{Relative differences of the activation rates of \kqu, \th, and \u, with respect to the ones evaluated in lowest irradiation position of the LS--25 channel.}
\label{fig:RelDiff_R}
\end{figure}

To analyze the systematic uncertainty due to the vertical flux profile, we calculated the relative differences of the \kqu, \th, and \u activation rates with respect to the bottom position in the Lazy Susan facility.
The results, shown in Fig.~\ref{fig:RelDiff_R}, are characterized by decreasing trends (similar to those measured with Au and Co samples) in which the maximum variation of the activation rate is $\sim9\%$ for \kqu, $\sim12\%$ for \th and up to $\sim18\%$ for \u.

\begin{figure*}[htb]
\centering
\begin{subfigure}{0.48\textwidth}
    \includegraphics[width=\textwidth]{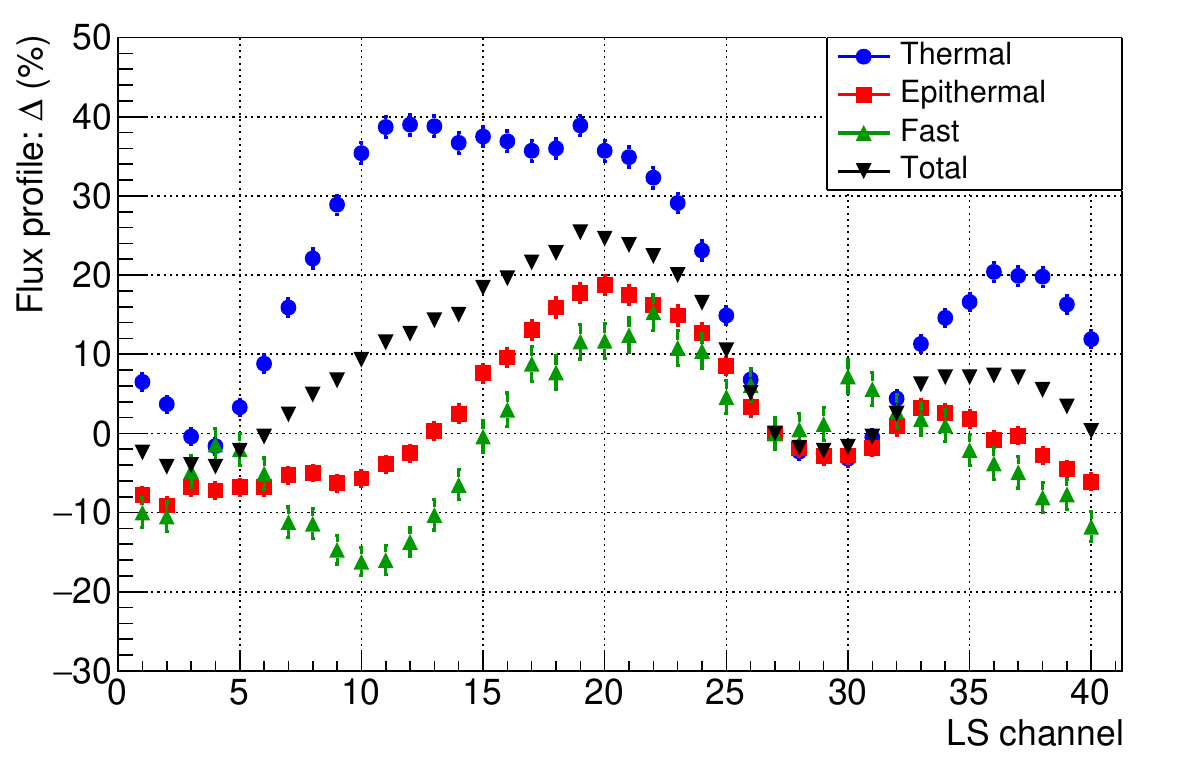}
    \caption{ }
    \label{fig:Flux_CircularMap}
\end{subfigure}
\hfill
\begin{subfigure}{0.48\textwidth}
    \includegraphics[width=\textwidth]{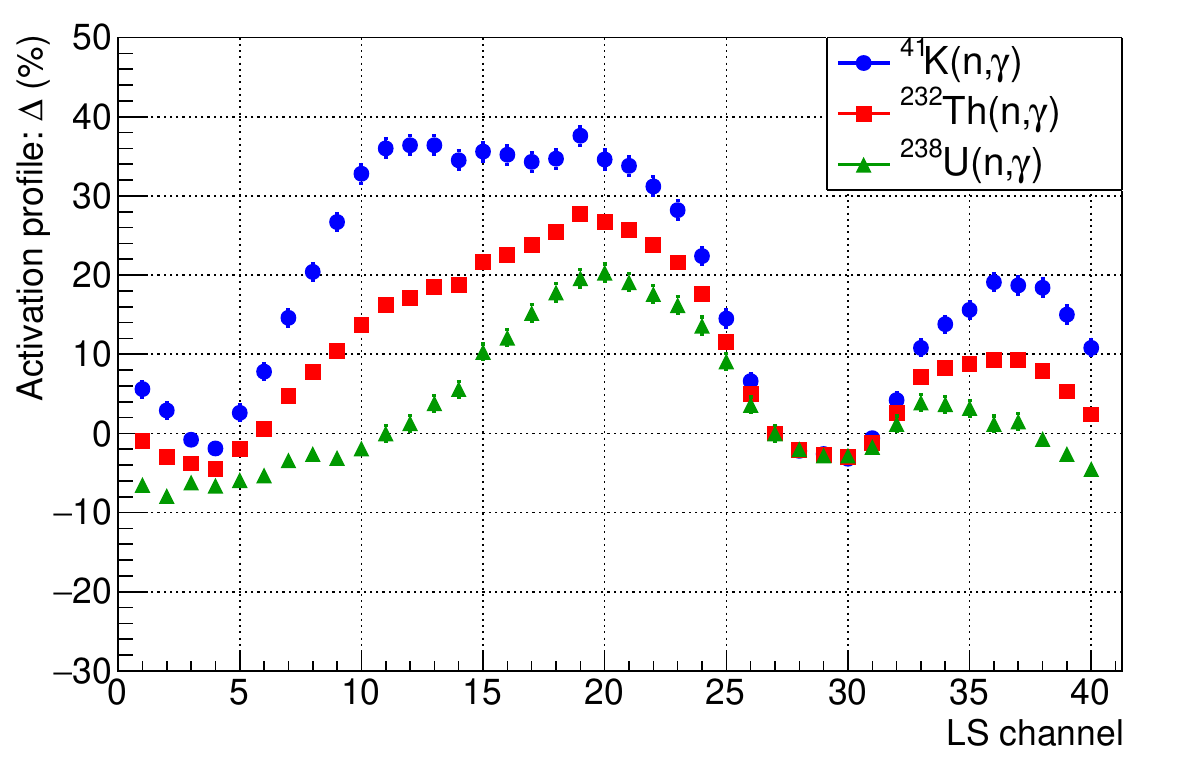}
    \caption{ }
    \label{fig:ASS_CircularMap}
\end{subfigure}
\caption{(a) MCNP simulation results of neutron flux in the annulus that houses the 40 channels of the Lazy Susan facility. The plot shows the flux relative difference with respect to the reference channel LS--27. (b) Activation rates of $(n,\gamma)$ reactions on \kqu, \th, and \u calculated in the different LS channels by combining MCNP flux results and group effective cross sections reported in Table~\ref{tab:XS_eff}.}
\end{figure*}

The results obtained so far allowed us to explore the impact of two possible sources of bias in activation measurements: the presence of a moderator material and the flux gradient along the vertical axis. 
The last source of systematic uncertainty that we investigated in this work is related to the flux variations in the annulus that houses the 40 channels of the Lazy Susan facility. 
In fact, this facility can no longer be rotated during irradiations, and the geometric asymmetry of the core and reflector causes flux variations.

In the absence of dedicated NAA measurements, we exploit the MCNP model of the reactor which, although not perfect, can give us an approximate estimate of the flux variations in the Lazy Susan annulus.
In Fig.~\ref{fig:Flux_CircularMap} we show the thermal, epithermal, fast, and integral flux profiles (normalized to 1 in correspondence with the reference channel LS--27) obtained by tallying the neutron flux in the bottom position of all 40 channels. 
According to the MCNP simulation model, the thermal flux can vary by up to $\sim40\%$ (epithermal and fast flux by up to $\sim30\%$) if channels far apart are considered. However the maximum difference between adjacent channels is $\sim7\%$ for all three flux groups and $\sim5\%$ for the total flux.
Finally, we combine these simulated fluxes with the group effective cross sections listed in Table~\ref{tab:XS_eff} to get an estimate of the activation rate profiles of \kqu, \th and \u. In Fig.~\ref{fig:ASS_CircularMap} we show the results of this calculation, from which it is possible to see that there are regions characterized by higher or lower activation gradients. The maximum difference between adjacent channels is obtained around the position LS--25 and is $\sim7\%$ for \kqu, $\sim6\%$ for \th, and $\sim5\%$ for \u.

\section{Conclusions}
In this paper we summarized the key aspects of neutron activation methodology for material screening in rare event physics experiments and we presented the workflow we developed at the Milano-Bicocca Radioactivity Laboratory to analyze samples irradiated at the TRIGA reactor of Pavia.

We paid special attention to the investigation of systematic uncertainties that may affect experimental results. In particular, through dedicated measurements, we assessed the effect of a moderator material surrounding the sample and we characterized the activation gradient in a typical irradiation configuration. 
By analyzing three different activation reactions we were able to unfold the thermal, epithermal and fast flux components in the different irradiation configurations.
Thanks to the MCNP simulation model of the TRIGA reactor, that we used to produce a set of template spectra for the unfolding analysis of activation data, we determined the energy spectrum of the neutron flux in the Lazy Susan irradiation facility with good accuracy. 
The detailed characterization of the intensity, spatial gradient and energy spectrum of the neutron flux allows us to calculate the effective cross section and the activation rate of any reaction of interest in different irradiation conditions. In this way it is possible to assess the systematic bias that can affect a neutron activation measurement. The bias varies depending on the reaction under investigation and on the geometry/material of the irradiated samples.

The sensitivities on \k, \th and \u contaminations that we can potentially achieve in the absence of interfering activation products is $<0.01$\,ppt for \k ($<0.1$\,ppb for elemental K), $<1$\,ppt for \u and of the order of a few ppt for \th contaminations. 
Based on our experience, it is quite common to activate interfering nuclides that worsen the sensitivity in the search for natural radionuclides. Therefore, when possible, we employ radiochemical sample treatments and background reduction techniques based on $\beta - \gamma$ and $\gamma - \gamma$ coincidence measurements, which can further improve our sensitivity and will be the subject of future articles.

\section*{Acknowledgments}
We thank the staff of the Laboratorio Energia Nucleare Applicata of the University of Pavia for the support provided to us to carry out neutron irradiations.
This work makes use of the Arby software for GEANT4 based Monte Carlo simulations, that has been developed in the framework of the Milano-Bicocca R\&D activities and that is maintained by O. Cremonesi and S. Pozzi.

\section*{Data availability statement}
The authors declare that the data supporting the findings of this study are available within the paper. The numerical data used to produce the figures are available from the corresponding author on reasonable request.

%
\bibliographystyle{spphys} 
\bibliography{Bibliography}

\end{document}